\documentclass[12pt,tightenlines,aps]{revtex4}

\usepackage{amsmath}
\usepackage{color}
\usepackage{setspace}
\usepackage{graphicx}
\usepackage{wrapfig}
\usepackage{subfigure}
\usepackage{epsfig}

\newcommand{\mike}[1]{{\color{black} #1}}

\usepackage[normalem]{ulem}

\begin{document}

\title{Avalanches mediate crystallization in a hard-sphere glass}

\author{E. Sanz$^1$, C. Valeriani$^1$, E. Zaccarelli$^2$, W. C. K. Poon$^3$, M. E. Cates$^3$ and P. N. Pusey$^3$}
\affiliation{$^1$Departamento de Qu\'{\i}mica F\'{\i}sica,
Facultad de Ciencias Qu\'{\i}micas, Universidad Complutense de Madrid,
28040 Madrid, Spain\\
$^2$CNR-ISC and Dipartimento di Fisica, Universit\`a di Roma La Sapienza, P.le A. Moro  2, 00185 Roma, Italy\\
$^3$SUPA, School of Physics and Astronomy, University of Edinburgh, Mayfield Road, Edinburgh, EH9 3JZ, Scotland}

\maketitle

%\begin{article}

%1- justify the sigma/3

{\bf  By molecular dynamics simulations we have studied the
devitrification (or crystallization) of aged hard-sphere glasses. First we find
that the dynamics of the particles are intermittent: quiescent periods, when
the particles simply ``rattle" in their nearest-neighbor cages, are interrupted
by abrupt ``avalanches" where a subset of particles undergo large
rearrangements. Second, we find that crystallization is associated with these
avalanches but that the connection is not straightforward. The amount of
crystal in the system increases during an avalanche but most of the particles
that become crystalline are different from those involved in the avalanche.
Third, the occurrence of the avalanches is a largely stochastic process.
Randomizing the velocities of the particles at any time during the simulation
leads to a different subsequent series of avalanches. \mike{The spatial
distribution of avalanching particles appears random, although correlations are
found among avalanche initiation events.} By contrast, we find that
crystallization tends to take place in regions that already show incipient
local order. }

Glasses are formed from the supercooled liquid state when motion is
arrested on the scale of the particle diameter. Such states are
thermodynamically unstable and may crystallize during, or shortly after, the
initial quench. (This is the usual fate of so-called ``poor" glass formers.)

Computer simulations have shown that in such cases crystallization readily proceeds 
by a sequence of stochastic micronucleation events that enhance the mobility in 
neighbouring areas, leading to a positive feedback for further crystallization \cite{sanzPRL}. 
%Computer simulations have shown that in such cases crystallization proceeds by
%a sequence of stochastic micronucleation events, correlated in space by dynamic
%heterogeneity \cite{sanzPRL}. 
Importantly however, crystallization can also
arise in mature, well-formed glasses after a long period of apparent stability.
The microscopic mechanism of this process, known as ``devitrification'',
remains elusive. Here we simulate the dynamics of a mature hard-sphere glass,
and find that crystallization is associated with a series of discrete
avalanche-like events characterized by a spatiotemporal burst of particle
displacements on a sub-diameter scale. The locations of these avalanches cannot be predicted from the prior
structure of the glass, and they vary among replicate runs that differ only in
initial particle velocities. Each avalanche leads to a sharp increase in
crystallinity, but remarkably the crystallizing particles are primarily not
those that participated in the avalanche itself. Instead they tend to lie in
nearby regions that are already partially ordered. We argue that a structural
propensity to crystallize in these regions is converted into actual
crystallinity by small random disturbances provided by the displacement
avalanche. Though spontaneous rather than externally imposed, this pathway may
relate to designed crystallization protocols such as oscillatory shear.

Devitrification is a phenomenon of both fundamental interest
\cite{zanottobook,kelton} and practical importance
\cite{metallic,schroers,schroers2,bulletin,opticalbook,ceramics,holand}.
Indeed the prediction and avoidance or control of devitrification represent
major formulation issues in materials science, arising for both metallic
\cite{metallic,schroers,schroers2} and network glasses
\cite{bulletin,opticalbook} as well as glass ceramics \cite{ceramics,holand}.
So far, however, there is limited understanding of the mechanisms whereby an
apparently deeply-arrested amorphous material can transform itself into a
crystalline packing without the large-scale, diffusive particle motions whose
absence (stemming from the formation of cages \cite {pusey1}) is a defining
property of glasses.

To gain such a mechanistic understanding, we study here by molecular dynamics
(MD) simulation what is probably the simplest model of a glass: a metastable,
amorphous assembly of equal-sized hard spheres in thermal motion. These systems
undergo a glass transition at a volume fraction of $\phi=\phi_g\simeq 0.585$
\cite{zaccarelli2009}. However, when the glass is prepared by rapid compression
to a density just above $\phi_g$, crystallites develop and grow almost
immediately \cite{sanzPRL,zaccarelli2009}. Put differently, monodisperse
glasses normally crystallize before reaching maturity, where we define
``maturity''  by persistence of the glass for decades beyond the molecular time\if{(hereafter referred to as `macroscopic' time scales)}\fi. This has so far
precluded using hard spheres as a model system for studying the devitrification
of a mature glass.

Recently however we have shown that mature monodisperse glasses can be created
by a numerical protocol called ``constrained aging''  \cite{constrained}, in
which motions that increase the global crystallinity are actively suppressed.
This protocol can be viewed as selecting only
the minority of dynamic trajectories in which the fresh (newly quenched) glass
accidentally outlives the quench.

In what follows, we present MD results for crystallization in these mature glasses at $\phi = 0.61$. This enables us to give a detailed mechanistic analysis of the devitrification process, in what is arguably the simplest model system available. 
We work at fixed volume \cite{sanzPRL,zaccarelli2009,constrained} 
%Following \cite{zaccarelli2009,sanzPRL,constrained} we work at fixed volume; this  is chosen 
to match the conditions in colloidal glasses, which are the nearest experimental realization of the hard-sphere model system and have long formed a key testing ground for glass physics concepts \cite{pusey1,weeksreview}.

Our first finding is
%We will first show 
that particle dynamics in a mature glass are intermittent:
quiescent periods of intra-cage motion are punctuated by `avalanches' in which a correlated subset of particles undergo cage-breaking displacements. Dynamic heterogeneities in glasses \cite{miyagawa,kobbarrat,schweizer,britowyart,cipelletti,katharina,yunker} (as opposed to supercooled liquids \cite{buchner,harrowellnatphys,harrowellavalanche,chandler1,onuki,appigna1}) have been reported previously, but avalanches have not been investigated in detail and no link has yet been made with crystallization dynamics. Importantly therefore, our second finding is that crystallization is intimately associated with these avalanches. This connection is however subtle: crystallinity increases during the avalanche, but most of the crystallizing particles are {\em not} among those taking part in the avalanche itself.
Thirdly, both the avalanche sequence and final crystallization pattern are stochastically determined: they depend not only on the initial particle coordinates but on their velocities, and change if these are re-assigned (following \cite{harrowellrandom}) in mid-simulation. 
Finally, we nevertheless find that crystallization preferentially occurs
in regions already showing semi-crystalline correlations or `medium range crystalline order' (MRCO) \cite{tanaka1,tanaka2,tanaka3}.

While certain of the above features can be individually discerned in our previous study of crystallization in fresh glasses \cite{sanzPRL}, only for mature glasses, which evolve more slowly, is the chain of causality between these events resolvable. 
%It is quite possible, however, that closely related mechanisms operate in fresh as well as mature glasses.

\section{Results}\label{sec:results}

\subsection{Avalanches}

Using the constrained aging method \cite{constrained}
we  generated a mature monodisperse hard-sphere glass of $\phi = 0.61$.  
This had an initially low
crystallinity, $X(0) \approx$~1\%, where crystallinity $X(t)$ is defined as the fraction of solid-like particles (the latter identified as described in {\it Methods}).
Starting from replicated initial particle coordinates,
we launched 15 MD runs, each having a different random (Maxwellian) set of particle momenta.
We have repeated the procedure for  different starting configurations, all producing similar results.
%FIGURE 1 A AND B
\begin{figure}[h!]
\begin{center}
\includegraphics[width=1.0\textwidth,clip=]{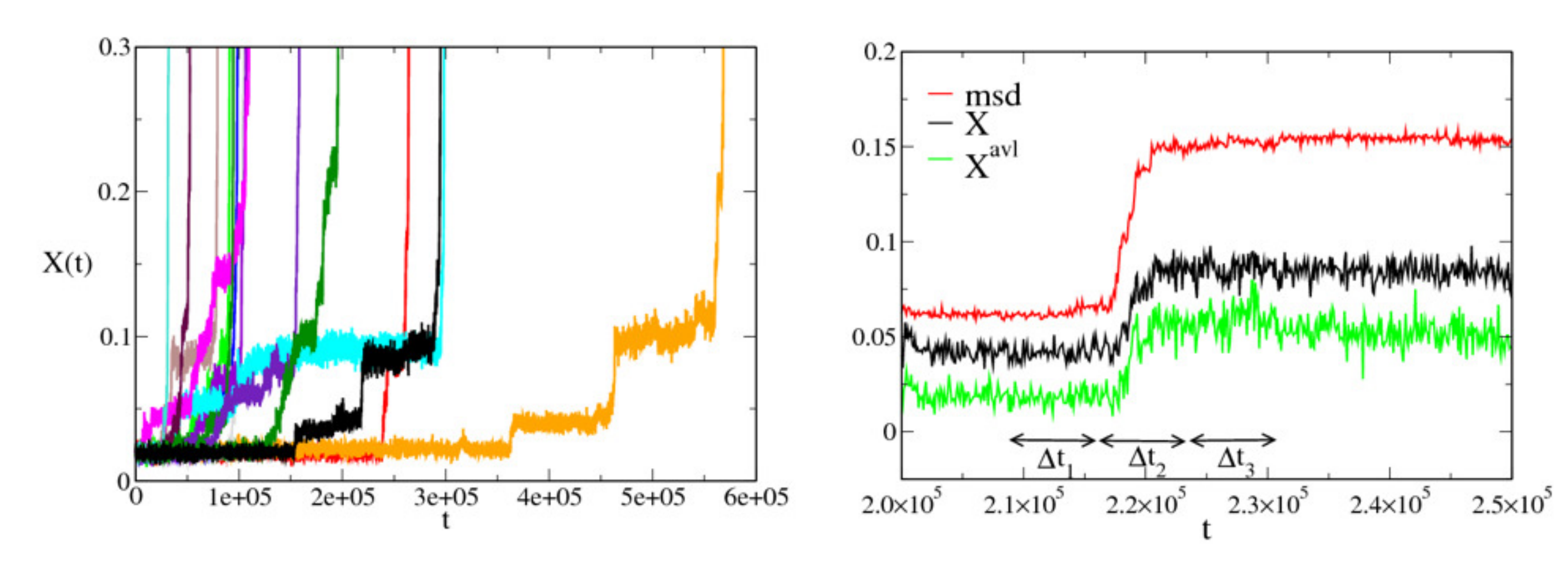}
\end{center}
\caption{({\bf a}) Fraction of solid-like particles $X(t)$ versus time for a system
of equal-sized hard spheres at volume fraction $\phi = 0.61$. 15 trajectories
are started from the same spatial configuration of particles but with different randomized momenta. ({\bf b}) Crystallinity $X$ (in black) and mean-square displacement (in red) versus time around the step-like
crystallization event shown in the black curve of Fig.~\ref{fig:figure1} at $t \approx 2.2 \cdot 10^5$.
The green curve, $X^{avl}$, is the fraction of avalanche particles defined in time interval $\Delta t_2$ that are solid-like. 
\label{fig:figure1}}
\end{figure}

In Fig.~1a we show the growth of crystallinity $X(t)$ for these 15 trajectories.
One might expect that, since crystallization in a glass takes place with
only small (sub-diameter) particle motions~\cite{sanzPRL}, its course should depend only on the starting configuration of the particles and not on their velocities. 
Yet Fig.~1a shows that the 15 replicas
have strongly dissimilar $X(t)$ profiles. This establishes a key role for stochasticity in the devitrification of mature glasses, like that reported previously for the crystallization of freshly formed ones~\cite{sanzPRL}. However, the $X(t)$ curves seen here for devitrification differ qualitatively from those of fresh glasses (Fig. 1a of \cite{sanzPRL}) which show slow monotonic growth from the beginning of the run.
By contrast, in the mature samples, $X(t)$ stays constant for between two and five decades of time (measured in microscopic units, see {\em Methods}) before steep upward jumps in $X(t)$ are seen. (These features depend on system size, as we discuss later.) Since the crystal is locally denser than the glass, each such upward step in $X(t)$ increases the free volume and speeds the approach of the next step. Under this feedback, the system finally crystallizes catastrophically and $X(t)$ goes rapidly to 1.

Key mechanistic insights are gained when we analyze one of these step-like crystallization events in more detail.
The black curve in Fig.~1b is a close-up of the crystallinity jump shown
in the black curve of Fig.~1a at $t \approx 2.2\times 10^5$.
The mean-square displacement (MSD) (see {\it Methods}), is also
plotted (red curve).
First, we notice that $X(t)$ and the MSD are
strongly correlated: both quantities jump simultaneously.
To understand the MSD jump, we compute
displacement vectors ${\bf u}$ of individual particles over chosen time intervals $\Delta t$,
and select those with $|{\bf u}| > \sigma/3$, with $\sigma$ the particle
diameter $\sigma$. (This threshold is justified in the Supporting Information (SI) Appendix.)

Figure~2 shows these vectors as red arrows for
the time windows indicated in Fig.~1b.
In window $\Delta t_1$ the system is largely immobile; most particles rattle
locally in their cages and less than 1\%
undergo significant displacements.
During window $\Delta t_2$, which spans the jump, a burst of displacements is recorded, with around 25\% of all particles moving more than $\sigma/3$. After the jump (window $\Delta t_3$) the system returns to quiescence with
again less than 1\% of all particles moving significantly.
We call such a sequence an ``avalanche'' and denote those particles that move by more than $\sigma/3$ during the jump ``avalanche particles''
(see SI Appendix for a justification of this cut-off alongside a more quantitative statistical analysis of the avanches). 
It is clear from the red arrows in the second frame of Fig.~2 that these particles are not homogeneously distributed,  
but cluster into ``avalanche regions'', resembling in exaggerated form the milder
dynamic heterogeneities often reported on the fluid side of the glass transition \cite{kob_dh,ediger_dh,book_dh}.

%FIGURE 2
\begin{figure}[h!]
\centering
\includegraphics[width=0.45\textwidth]{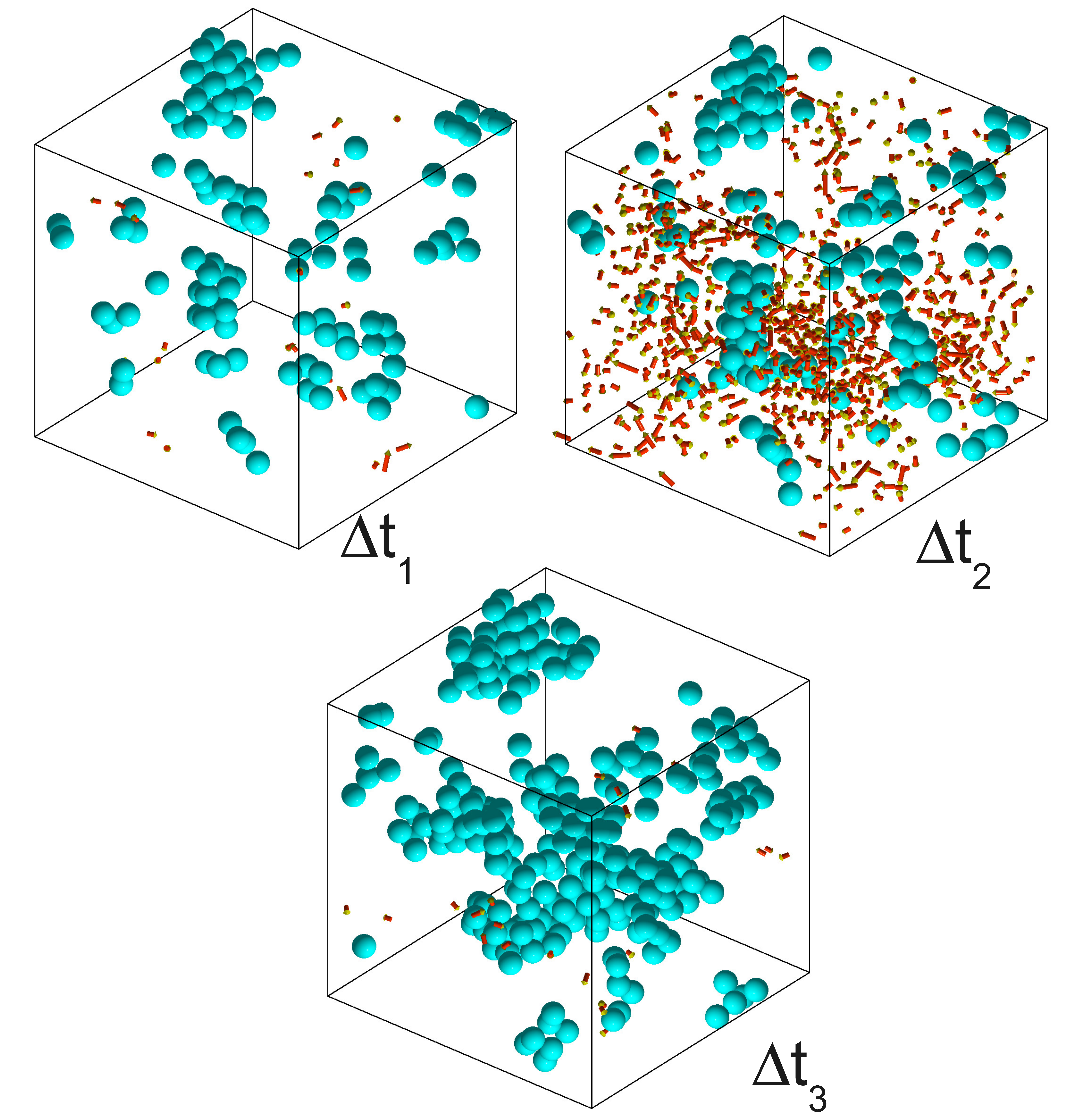}
\caption{\label{fig:avalanche2}
Displacement vectors with modulus larger than $\sigma/3$ (red arrows with yellow heads) and solid-like particles (turquoise spheres) 
for time intervals $\Delta t_1$, $\Delta t_2$ and $\Delta t_3$ shown in Fig.\ref{fig:figure1}b. The lengths of the arrows correspond
to the modulus of the displacements. Solid-like
particles are defined at the beginning of each time interval. }
\end{figure}

By interrogating the dynamics across narrower time intervals, we have observed
that avalanches start to build in localized regions, then grow to peak
activity, and finally die out (see online video in SI Appendix).
From start to finish, an avalanche typically takes about $7\times 10^3$ time
units. Highly cooperative movements can be seen during the main avalanche
phase, including particles moving in rows or circles (see Fig. 3).  Turquoise
spheres in Fig. 2 correspond to solid-like particles.  As expected from
Fig.~1b, the avalanche leaves behind an increased population of solid-like
particles.

%FIGURE 3
\begin{figure}[h!]
\centering
\includegraphics[width=0.5\textwidth]{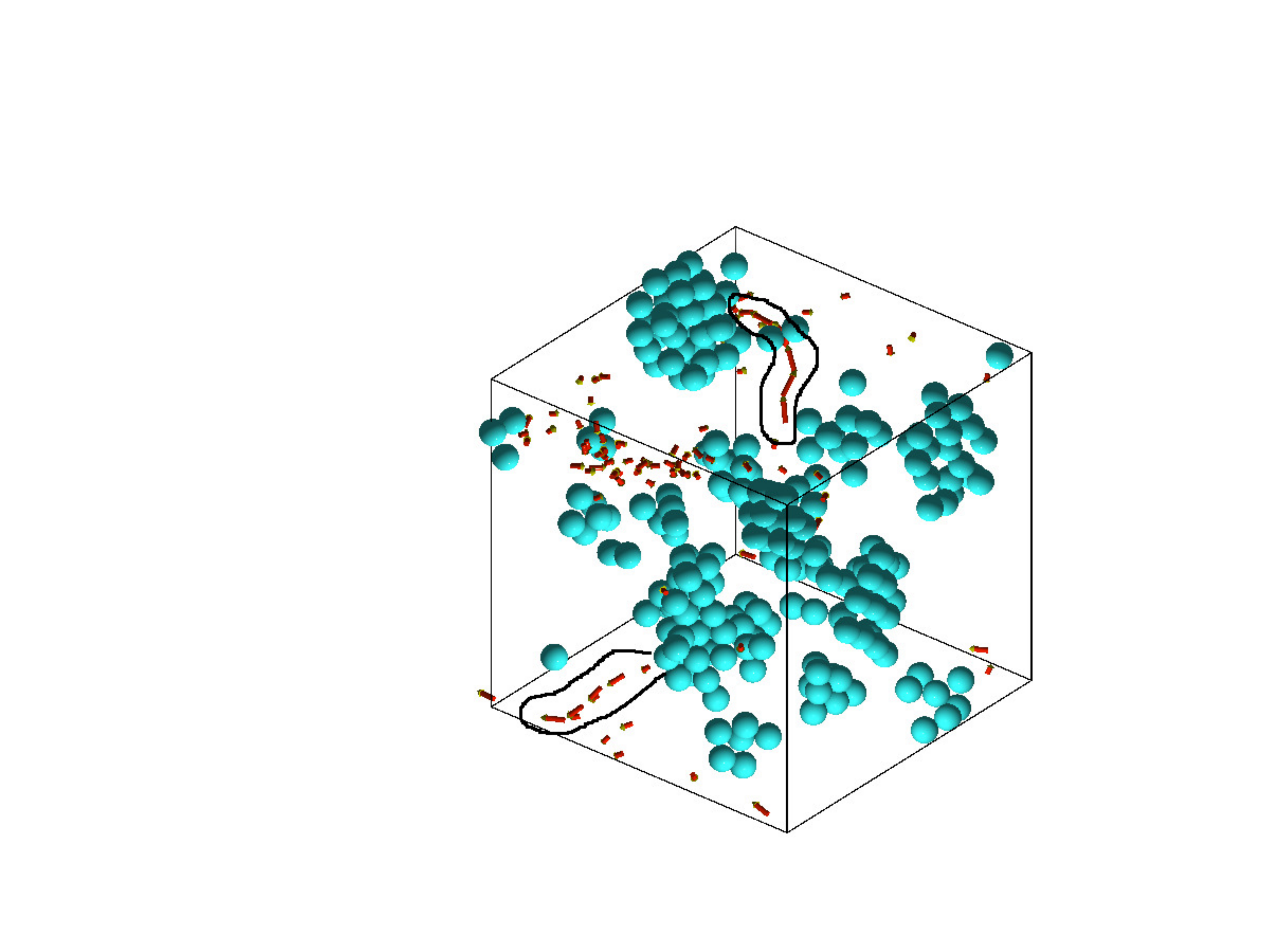}
\caption{\label{cooperative}
Displacement field for a typical avalanche in which cooperative motion where particles follow each other are highlighted.}
\end{figure}

\subsection{Avalanches mediate crystallization}

Figures~1b and 2 show one representative example of a jump in crystallinity
partnered with a displacement avalanche. This is a general phenomenon: in none
of the runs do we see crystallinity jumps that are not associated with
avalanches.  The question thus arises: do avalanches cause crystallization, or
vice versa?  If avalanches cause crystallization, one obvious hypothesis is
that the particles that move to become crystalline 
%become crystalline, in moving to do so, 
are the ones that form the avalanche.  However, this hypothesis can be ruled
out by visually inspecting Fig.~2 and realising that there is no clear overlap
between avalanche regions and regions where new crystalline particles appear.
The fraction of crystalline particles is $\simeq$4\% before the avalanche and
$\simeq$9\% afterwards.  Of the new crystalline particles, only 25\%  were
directly involved in the avalanche, as one can infer from the green curve in
Fig.~1b. (The proportion depends somewhat on the exact threshold of
displacement used to define avalanche particles.) We conclude that the
particles that crystallize are {\em mainly not} the ones that \mike{participated in} the
avalanche.

An alternative hypothesis is that avalanches are caused by crystallization in
the sense of being triggered by the small rearrangements ($|{\bf u}| <
\sigma/3$) \cite{zaccarelli2009,constrained} needed to achieve local
crystallinity. If so, avalanches would be absent whenever crystallization is
suppressed by size polydispersity.  

%FIGURE 4
\begin{figure}
\centering
\includegraphics[width=0.45\textwidth]{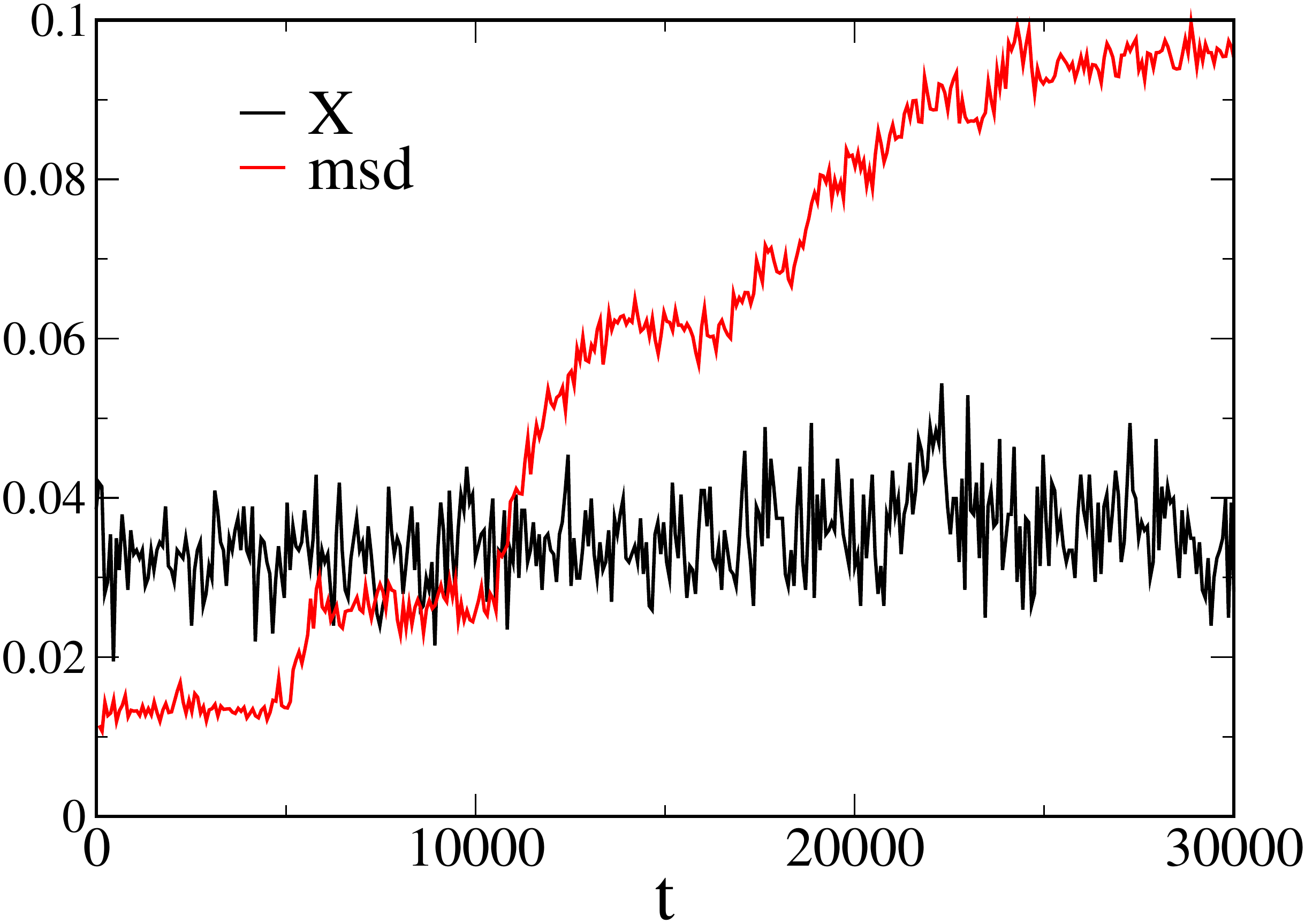}
\includegraphics[width=0.45\textwidth]{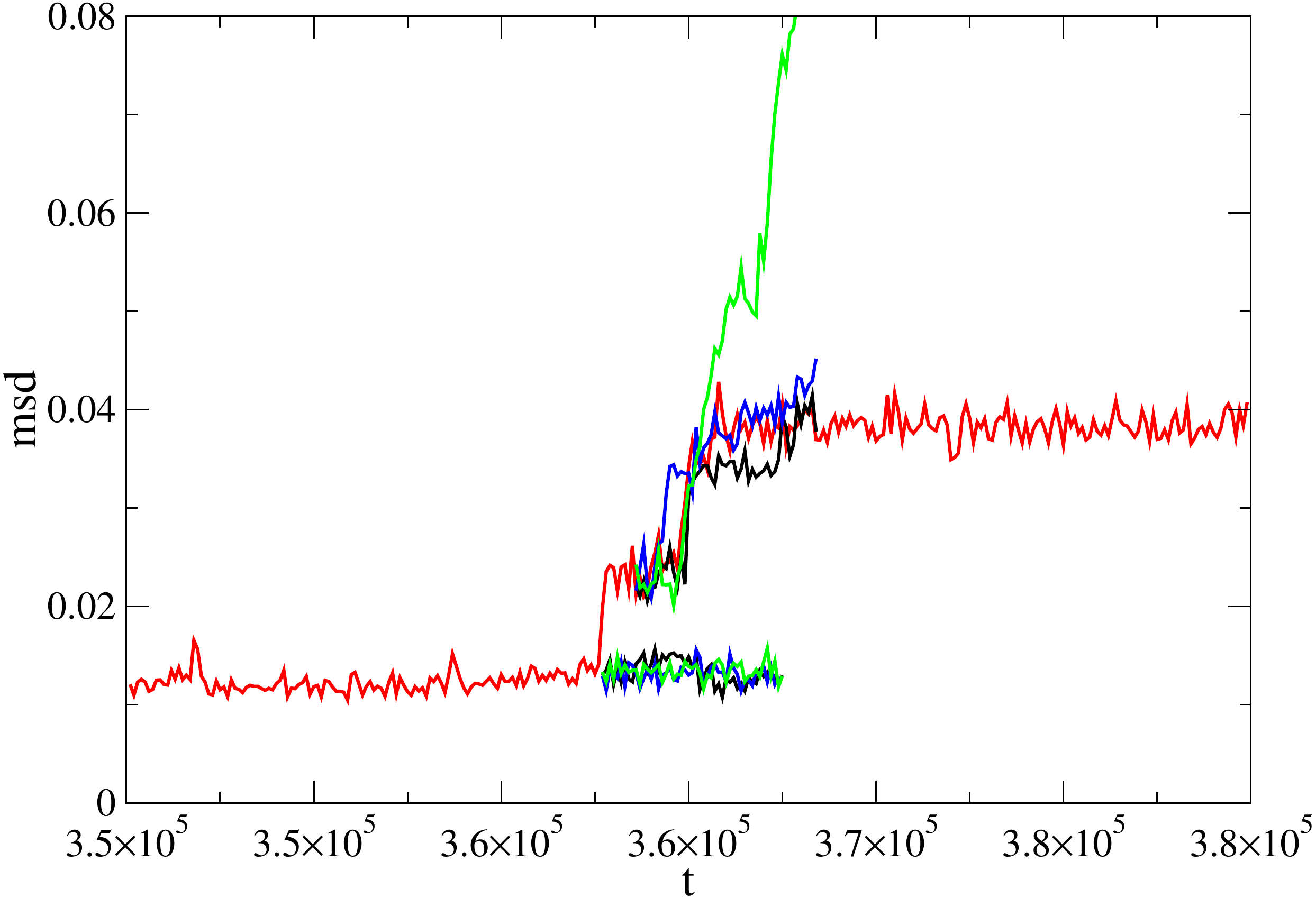}
\caption{\label{fig:avaglass1} ({\bf a}) Crystallinity $X(t)$ (black) and mean-square displacement (red) versus time  for a 6\%
polydisperse system at $\phi=0.60$. ({\bf b}) Red curve: mean-square displacement versus time for a trajectory of the
monodisperse system showing an avalanche.  Blue, green and black curves: MSDs for
the same system when the particle velocities are randomised immediately before the avalanche and in the middle 
of the avalanche.} 
\end{figure}

Figure~4a shows the MSD and $X(t)$ of a
glass with 6\% polydispersity at volume fraction $\phi=0.60$.  As expected from
our earlier work \cite{zaccarelli2009,puseyphiltrans}, the crystallinity stays
flat throughout the run; yet we see that the MSD jumps in a way that, by
the methods already described, can be identified as avalanches.  Moreover,
avalanche-like dynamic heterogeneity (in less extreme form) was previously seen
for other non-crystallizing glassy systems in 2D and 3D simulations
\cite{kobbarrat,britowyart,cipelletti,buchner,harrowellavalanche,chandler1} and
in colloid experiments \cite{cipelletti2,weeks2}.  Therefore we can discard the
hypothesis that crystallization causes avalanches, rather than vice versa.

The stochastic nature of avalanches was already shown in Fig.~1, where the
trajectory of each replica has a different crystallinity evolution $X(t)$.  A
further illustration is given in Fig.~4b, where we compare a trajectory
undergoing an avalanche with  three  systems started from a common
configuration just prior to the avalanche. Each replica is launched with a
different set of particle velocities, and in all three cases the avalanche is
averted.  This shows that the triggering of an avalanche from the quiescent
state does not depend on particle coordinates alone, but rather on the
appearance of a successful combination of positions and momenta. 
%On the contrary, when launching three trajectories with a different set of
%particle velocities from the middle of an avalanche
%(Fig.~\ref{fig:avaglass1}b),  particles are committed to create an avalanche,
%that cannot be averted. 
We speculate that these rare events involve emergence of cooperative motions
such as those illustrated in Fig.~3.  In contrast, if velocities are reassigned
midway through an avalanche (see Fig.~4b), the avalanche does not stop, but
continues along an altered path. This implies that the `activated' state is
structurally distinguishable from the quiescent one, although we have not yet
found a clear static signature for it.

The requirement of an unlikely combination of positions and velocities to
trigger an avalanche, combined with the fact that avalanches cause
crystallinity to grow (explored further below) explains the stochasticity of
devitrification in our mature samples and is likely also implicated in the
stochastic crystallization in fresh glasses \cite{sanzPRL}.  That displacement
avalanches mediate crystallization in hard-sphere glasses is the central
finding of this paper.

\subsection{Heterogeneities}
\begin{figure} 
\centering
\includegraphics[width=0.5\textwidth]{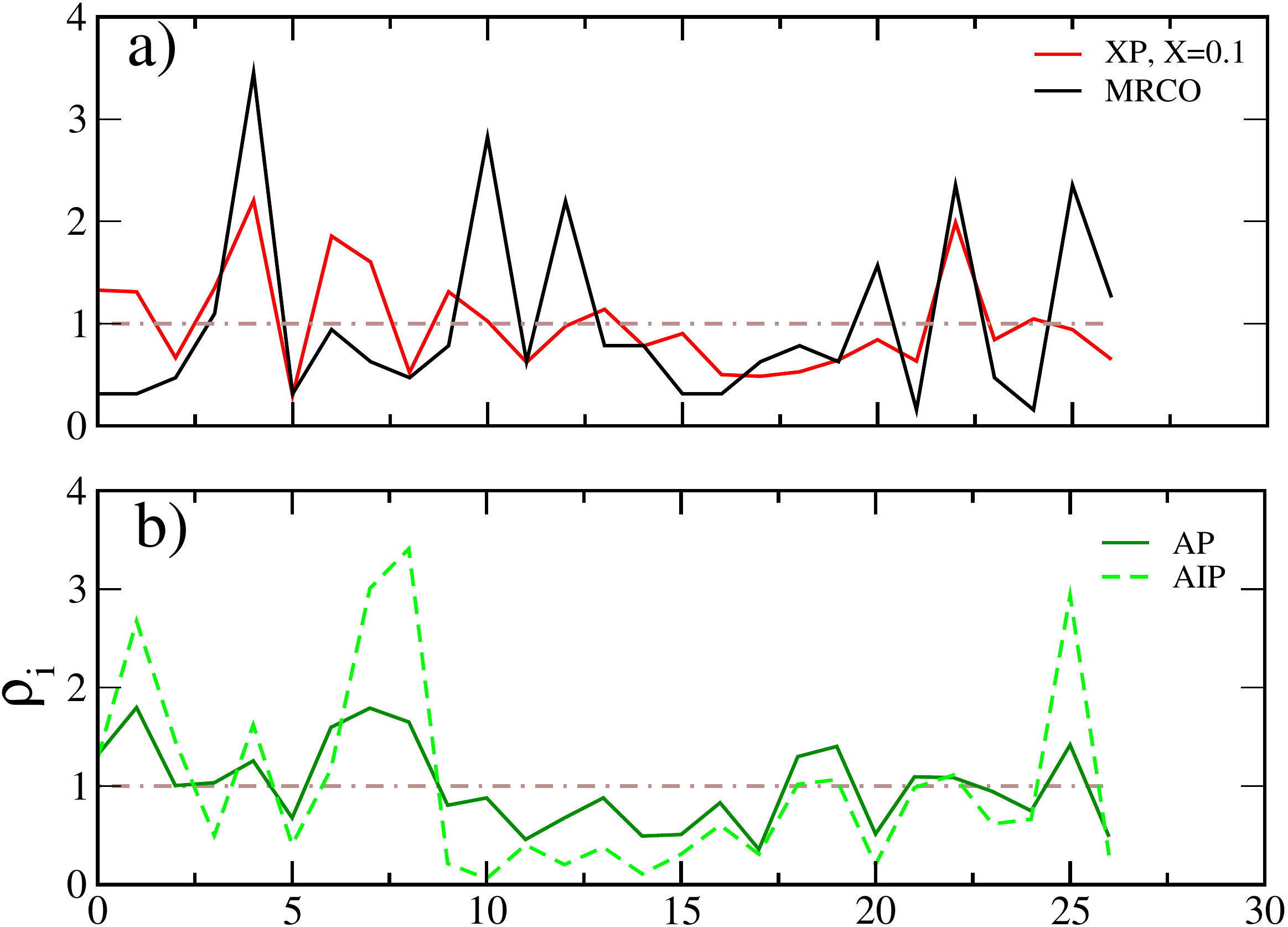}
\caption{\label{3propensities} Normalised density, $\rho_i$, as a function of
the index $i$ identifying each subvolume of the simulation box for various
particle types (see text).  
({\bf a})
Red: Crystalline particles (XP) are those of all the trajectories in Fig.
\ref{fig:figure1}a as they first cross the crystallinity threshold X = 0.1.
Black: Medium range crystalline order (MRCO) particles are those in the initial
configuration with bond order parameter $\bar{q}_6$ in the top 10\%. 
({\bf b}) 
Dark Green (solid line): Avalanche particles (AP)
are those \mike{participating} in the first avalanche of all trajectories.  
Light Green (dashed line): Avalanche initiator
particles (AIP) are those \mike{involved in initiating} the first avalanche of
all trajectories \mike{as defined in Supplementary Information.}}
\end{figure}

As previously stated, the different trajectories in Fig. 1 lead to different
final crystallization patterns from the same initial configuration. Visual
inspection of these patterns shows only limited similarity between them.
Nonetheless, one might expect some regions to be more likely to crystallize
than others. The crystallization propensity is assessed by superimposing the
crystalline particles (XP) of all trajectories as these first cross a fixed crystallinity
threshold (we choose $X = 0.1$).  To quantify any heterogeneity in the
resulting superimposition we divide the simulation box in 3x3x3 equal
subvolumes and evaluate the density in each, normalizing by the overall density. 
The resulting normalised densities, $\hat{\rho}_i=\rho_i/\rho$, are plotted as
a function of subvolume index in Fig.~\ref{3propensities}a.  
By computing the
fluctuations of $\hat{\rho}_i$ around the average value, 1, we get a
quantitative measure of the degree of heterogeneity, $h_d= \langle
\hat{\rho}_i^2 \rangle - \langle \hat{\rho}_i \rangle ^2$.  For crystalline
particles in our replicated runs we find $h_d = 0.22$, more than four times
above the background level, $h_d = 0.050$, computed by superposing
crystalline particles for 15 runs starting from {\em independent} initial
configurations rather than replicated ones.
We can conclude 
that there are some regions in the initial configuration that are more prone to
crystallize than others.  
It has been found in supercooled liquids that these
regions correlate with a partial ordering known as medium-range crystalline
ordering, MRCO, which is quantified by an averaged local bond order parameter
$\bar{q}_6$ \cite{tanaka2,tanaka3,dellago}.

%FIGURE 5

Fig. 5a compares the density of XP
particles in our simulations with the density of MRCO, identified as those
particles with $\bar{q}_6$ in the top 10\%.  As with the earlier work on
supercooled liquids \cite{tanaka2,tanaka3}, there is a clear, though not
complete, correlation between medium-range crystalline order in the initial
configuration and subsequent crystallization.

We also investigate if there are regions where avalanches have a higher propensity 
to take place by doing a similar analysis as that described above but for 
particles involved in the first avalanche (AP) instead.  
As seen in Fig. 5b, the density of these particles shows only small variations
between sub-volumes, suggesting that avalanches occur almost at random
throughout the system in  mature glasses (whereas the crystallinity induced by
these avalanches has a significantly higher propensity to appear 
in some regions than in others).

It has been found that dynamic heterogeneities in supercooled fluids, involving large-scale
rearrangement of the particle positions (the alpha process), 
tend to grow from regions 
of high displacement in low-frequency quasilocalized phonon modes (so-called soft spots)
\cite{britowyart,harrowellnatphys,harrowell2006,harrowell2009,yodh}.  
In view of our result that avalanches occur almost at random throughout the system
one would be tempted to conclude that avalanches and dynamic heterogeneities 
are fundamentally different dynamic events.  However, a closer study does
reveal a clear correlation across trajectories among avalanche {\em initiator}
particles (AIP, those involved in the first steps of avalanche formation). 
In fact, the density heterogeneities plot of AIP shown in Fig. 5b shows
large density variations between sub-boxes (in the SI Appendix we show that this is a 
statistically significant result). 
Therefore, AIP and dynamic heterogeneities share the tendency to 
develop in certain regions of the system. Whether or not these regions
also correspond to soft spots for the case of mature glasses requires further
investigation beyond the scope of this paper. 
Nevertheless, we show some 
preliminar analyses in the SI Appendix, alongside  
a more detailed account of heterogeneities, including pictorial representations.

\section{Discussion and Conclusions} 

We have investigated the mechanism by which crystals develop in amorphous
glasses composed of equal-sized hard spheres.  In contrast with our previous
work on freshly prepared samples, we addressed here mature glasses, whose
arrest is characterized by a mean-square displacement that stays flat for
several decades in time prior to the onset of crystallization.  We have shown
that crystallization is intimately associated with particle displacement
avalanches (Figs.~1b and 2), and that crystallization is caused by these
avalanches and not vice versa. However, the majority of \mike{avalanche participants} do not become crystalline (green curve in Fig.~1b), and most
crystallizing particles move little 
during the avalanche.  Thus the displacement avalanche is not, of itself, the
sequence of motions needed to transform an amorphous region into crystal.

\mike{Instead, avalanches within the mature glass appear to be autonomous structural
rearrangements, involving cooperative particle motion. These mesoscopic
avalanches have a strongly stochastic character, and are triggered by unlikely local
combinations of particle positions and momenta. An individual avalanche can be
averted entirely by reassigning momenta just before its inception; once
underway, however, such reassignment only diverts it along a different path
(Fig. 4). Although no obvious propensity to occur in particular positions can be seen in the statistics of avalanche participants, this can be detected among avalanche initiator particles. This implies a correlation with static structure (explored further in SI Appendix), possibly including `soft spots' of the type known to be linked to dynamic heterogeneity in supercooled liquids \cite{britowyart,harrowell2006,harrowell2009} and some glasses
\cite{britowyart,harrowellnatphys,yodh}.
If so, our avalanches might be viewed as a limiting type of dynamic heterogeneity, arising as the system's density or age increases so that activity becomes rare. However, the stochastic character of the avalanches might also be taken as support for suggestions \cite{britowyart,harrowellavalanche} that a qualitatively different type of dynamics takes over in systems, such
as ours, that are deep into the glassy state.
In addition, and in common with supercooled liquids, we} find that the crystals
tend to grow in regions of medium range crystalline order (which seem to be themselves
{\it anti}-correlated with the soft spots \cite{tanaka1,tanaka2,tanaka3}, see Fig. 8 of Supplementary Information).

The likely role of avalanches in crystallization is to create the small
disturbances required to accomplish ordering in regions that, as noted above,
already have a propensity to crystallize.  Avalanche-induced disturbances might
shake a nearly ordered region into order, but could also facilitate growth of
an established crystallite at its perimeter. This avalanche-mediated mechanism
for devitrification somewhat resembles the breakdown dynamics of an attractive
colloidal gel \cite{cipellettiball}. The process could also be closely related
to protocols such as shearing in which mature glasses are induced to
crystallize by gentle agitation \cite{vanmegen,haw,petekidis}. In contrast to
those protocols, here the required agitation is spontaneously generated. Indeed
the intrinsic avalanche dynamics remain present even when crystallization
itself is prevented by polydispersity.

In keeping with previous findings for fresh glasses
\cite{sanzPRL,zaccarelli2009}, the ordering induced by an avalanche reduces the
pressure in the system and creates positive feedback for further avalanches.
This gives rise to a nontrivial system size dependence for the time evolution of
global properties such as the mean crystallinity, as explained in the
SI Appendix. However it does not qualitatively change the mesoscopic
mechanism of avalanche-mediated devitrification that we have described.

%In order to confirm that our findings are not some special feature of systems prepared by constrained aging, we have additionally performed some simulations on fresh glasses prepared by rapid compression to an even higher concentration, $\phi = 0.62$.  We found that these glasses showed similar behavior to that reported above for the constrained-aged systems at $\phi = 0.61$: long quiescent periods and sudden coincident jumps in the crystallinity andmean-square displacement (see SI Appendix).  
In order to confirm that our findings are not some special
feature of systems prepared by constrained aging, we have
additionally performed simulations on fresh glasses prepared by rapid compression 
to a higher concentration, $\phi=0.62$, where there is no need to resort to constrained aging to 
obtain a mature glass. We found that these glasses show similar behavior 
to that reported above for the constrained-aged systems at $\phi=0.61$: 
long quiescent periods and sudden coincident
jumps in the crystallinity and mean-square displacement (see
SI Appendix). Therefore, this devitrification mechanism is evident for mature glasses, 
either prepared by constrained aging ($\phi=0.61$) or by 
quick compression ($\phi=0.62$). By contrast, a glass prepared by quick compression 
at $\phi=0.61$\cite{sanzPRL} crystallizes while still fresh and does not clearly show the avalanche mechanism.

Our work suggests several avenues for future research. One is to study hard
sphere devitrification at constant pressure. A second is to address by our
methods mixtures of different-sized hard spheres. This would represent a first
step towards modelling bulk metallic glasses, which are generally
multi-component alloys \cite{ruta,baldi}. Mechanistic insights along the lines pursued in this
paper might then shed light on the devitrification of such glasses during
processing, which is a major issue in technology  \cite {schroers,schroers2}. 

%An obvious extension of this work would be to mixtures of different-sized hard spheres, thus modelling bulk metallic glasses where devitrification is a major issue during the processing stage \cite {schroers,schroers2}. 

\section{Materials and Methods}

\subsection{Simulation details} \label{sec:methods}

We perform event-driven Molecular Dynamics simulations in the $NVT$ ensemble with cubic periodic
 boundary conditions for a system of $N$ = 3200 monodisperse hard spheres~\cite{rapaport,zacca1}.
We also simulate a polydisperse system of $N$ = 2000 particles 
where the particle diameters are chosen
according to a discrete Gaussian distribution with relative standard
deviation $s=0.06$.
Mass, length, and time are measured in units of particle mass $m$, particle diameter $\sigma$ (or 
$\bar{\sigma}$ for the polydisperse case)  and
$t_0=\sqrt{m \sigma^2/\kappa_B T}$, where $\kappa_B$ is the Boltzmann constant and $T$ the temperature,
and we set $\kappa_B T=1$.
The packing fraction is defined as $\phi=\frac{\pi}{6} N \sigma^3/V$ (with $V$ the system's volume).

To generate the initial configuration we follow the `constrained aging' procedure
described previously\cite{constrained}.
We use a configuration resulting from constrained aging
as a starting point for unconstrained Molecular Dynamics runs.

\subsection{Analysis details}

The mean-square displacement, MSD, is calculated as $\frac{1}{N} \sum_{i=1}^{N} (\mathbf{r}_i(t) - \mathbf{r}_i(0))^2$,
where  $\mathbf{r}_i$ is the position of particle $i$.

The crystallinity, $X$,
is defined as the number of solid-like particles divided by the total number of
particles.
As in  previous work\cite{puseyphiltrans}, we identify solid-like particles according  to a rotationally
invariant local bond order parameter $d_6$ \cite{d6_1,d6_2}.
In order to compute it, we first identify the number of neighbours $N_b(i)$ of each particle $i$ using the
parameter-free SANN algorithm \cite{koos}.
Next,  for every particle $i$ we compute the complex
vector $\mathbf{q}_{6}$ whose components are given by
$q_{6m}(i) = \frac{1}{N_b(i)} \sum_{j=1}^{N_b(i)} Y_{6m} (\theta_{ij},\phi_{ij}) / 
\left( \sum_{m=-6}^{6} q_{6m}(i) \cdot q^*_{6m}(i) \right)^{1/2}$ (with $m \in [-6,6]$)
where  $Y_{6m}$ are $6^{th}$ order spherical harmonics.
Then we compute the rotationally invariant bond order parameter $d_6$
by calculating the scalar product between each particle's $\mathbf{q}_{6}$
and its neighbours,
$d_6(i,j) = \sum_{m=-6}^{6} q_{6m}(i) \cdot q^*_{6m}(j)$, and
consider particles i and j as having a "solid connection" if their
$d_6(i,j)$ exceeds the value of 0.7.
A particle is labelled as solid-like if it has at least 6 solid
connections.

\begin{acknowledgments}
 CV and ES acknowledge financial support from  an Intra-European Marie Curie Fellowship (in Edinburgh, 
 237443-HINECOP-FP7-People-IEF-2008 and 237454-ACSELFASSEMBLY-FP7-People-IEF-2008
respectively) and from a Marie Curie Career Integration Grant (322326-COSAACFP7-
PEOPLE-CIG-2012 and 303941-ANISOKINEQ-FP7-PEOPLE-CIG-2011
respectively), together with the Juan de La Cierva (JCI-2010-06602) and Ramon y Cajal
(RYC-2010-06098) Spanish Fellowships and the Spanish National Project FIS2010-
16159. EZ acknowledges support from MIUR-FIRB ANISOFT ( RBFR125H0M).
WCKP, MEC and EZ acknowledge support from ITN-234810-COMPLOIDS and WCKP and MEC for the EPSRC grant  EP/J007404. MEC holds a Royal Society Research Professorship.
This work has made use of the resources provided by the Edinburgh Compute and Data Facility (ECDF). The ECDF is partially supported by the eDIKT initiative.
 \end{acknowledgments}

\newpage

\section{Supporting Information to ``Avalanches mediate crystallization in a hard-sphere glass''}

\subsection{Avalanche definition and statistics}\label{appava}

\begin{figure}[h!]
\begin{center}
\includegraphics[width=0.4\textwidth,clip=]{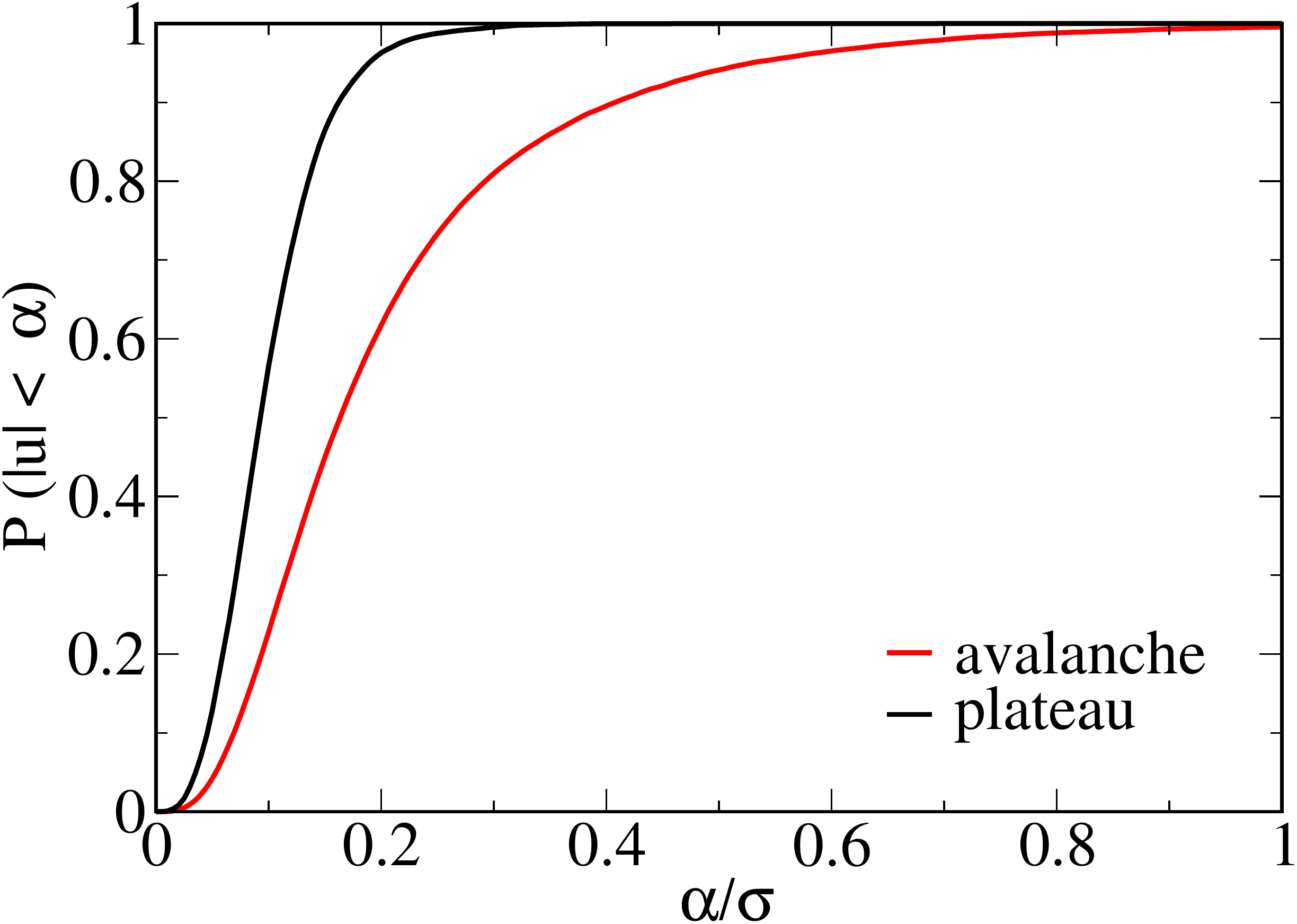}
\caption{
Probability that the displacement of a particle is smaller than a certain distance, $P(|{\mathbf u}|<\alpha)$, versus 
the distance, $\alpha$, in particle diameters. 
We compare $P(|{\mathbf u}|<\alpha)$ for a quiescent period (black) with $P(|{\mathbf u}|<\alpha)$ when 
an avalanche takes place (red). 
In the quiescent plateau, $P(|{\mathbf u}|<\alpha)$ is calculated for a time interval equal to the 
average duration of an avalanche ($7000t_0$). 
\label{fig:plateaxdisp}
}
\end{center}
\end{figure}

We define avalanche particles as those whose displacement ($|{\mathbf u}|$) during a given time interval 
is larger than $\sigma/3$. To show that such displacements are indeed significantly large we
compute the cumulative probability distribution of displacements 
in the initial quiescent plateau (before the first avalanche) for a time interval
equal to the average duration of an avalanche. 
By inspecting the 15 trajectories shown in Fig. 1 in the main text we found that 
an avalanche lasts on average about $7000 t_0$. (Here $t_0$ is the time unit introduced in the Methods section.)
The black curve in Fig. \ref{fig:plateaxdisp} represents $P(|{\mathbf u}|<\alpha)$, the probability that the displacement of a
particle is smaller than  $\alpha$, for a time interval of $7000 t_0$ in the initial quiescent plateau. 
Clearly, displacements larger than $\sigma/3$ are extremely rare 
in the quiescent period, which justifies our threshold for the definition 
of avalanche particles. By contrast, it is not unlikely that particles travel for even longer distances during an 
avalanche.  
This is demonstrated by the red curve in Fig. \ref{fig:plateaxdisp}, 
which corresponds to $P(|{\mathbf u}|<\alpha)$ calculated during a time interval 
that includes an avalanche.  
The curve is made with the collection of all the displacements during the first avalanche of each of the trajectories shown 
in Fig. 1 of the main text.
About 15 per cent of the particles travel more than $\sigma/3$ during an avalanche. 
This means that, according to our definition, an avalanche involves on average about 500 particles. 
Notice that particles do not move beyond their diameter during an avalanche and only 6 percent of 
them travel beyond the radius. Therefore, the mobility during an avalanche, even if 
much larger than that during a quiescent period, is still rather restricted.

%This appendix is included here for completeness.
%FIGURA 3

The rate at which avalanches nucleate is the limiting factor for the growth of crystals in a glass. 
We can estimate the avalanche nucleation rate for our configuration by counting the number of avalanches 
and dividing it by the time the system takes to fully crystallize and by the volume of the system. 
The value we get after averaging over all trajectories is 
6$\cdot 10^{-9}$  $\sigma^{-3} t_0^{-1}$. 
This nucleation rate implies that the first avalanche   
takes place, on average, in 6$\cdot 10^{4}t_0$ in our system of volume $14^3 \sigma^3$.
Of course, the larger the system's volume, the shorter the time 
it takes for the first avalanche to nucleate. 

Below we discuss the influence 
of the system size on the crystallization pathway, and present a more quantitative description of the avalanches than that 
given in the main text. This type of analysis should be interpreted with care, though.  
As discussed below, the qualitative picture of the crystallization mechanism 
is not affected by the way the configuration is generated. 
However, since we are dealing with a system out of equilibrium, the history 
of formation and, of course, the packing fraction,  
may have an impact over the precise value of the variables here discussed.

\subsection{Dependence on System Size}

\begin{figure}[h!]
\centering
\includegraphics[width=0.45\textwidth,clip=]{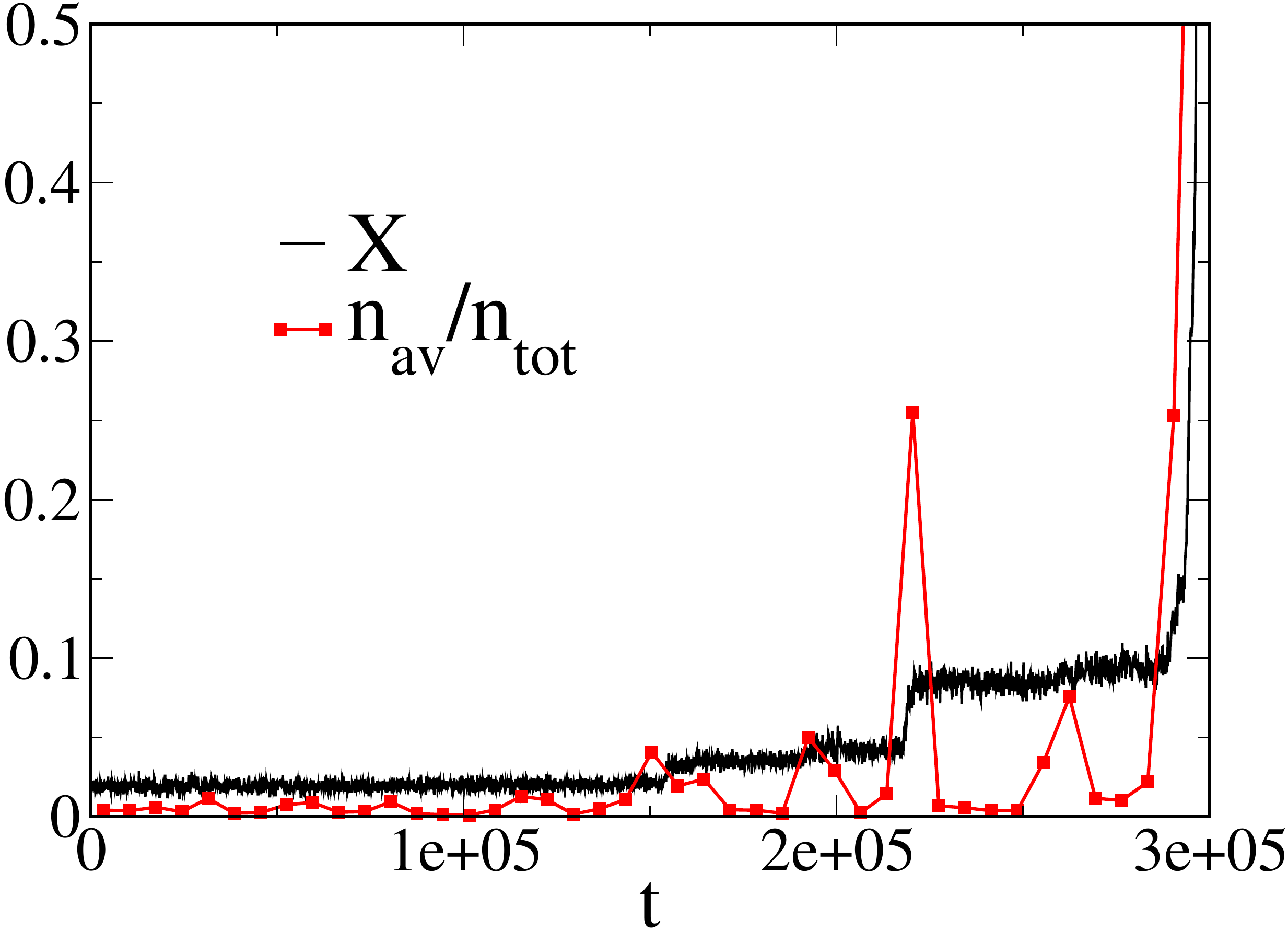}(A)
\includegraphics[width=0.45\textwidth,clip=]{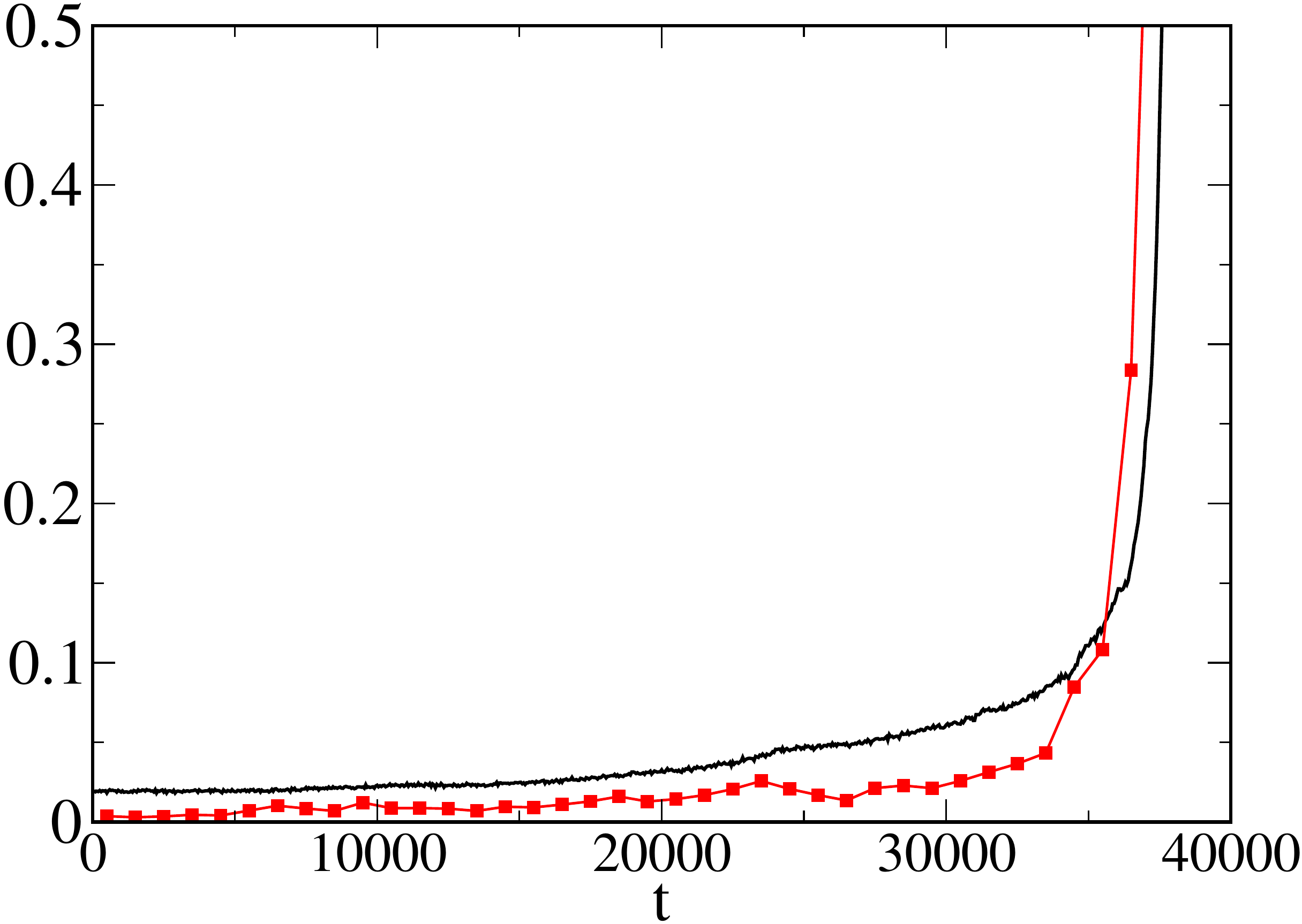}(B)\\
\includegraphics[width=0.45\textwidth,clip=]{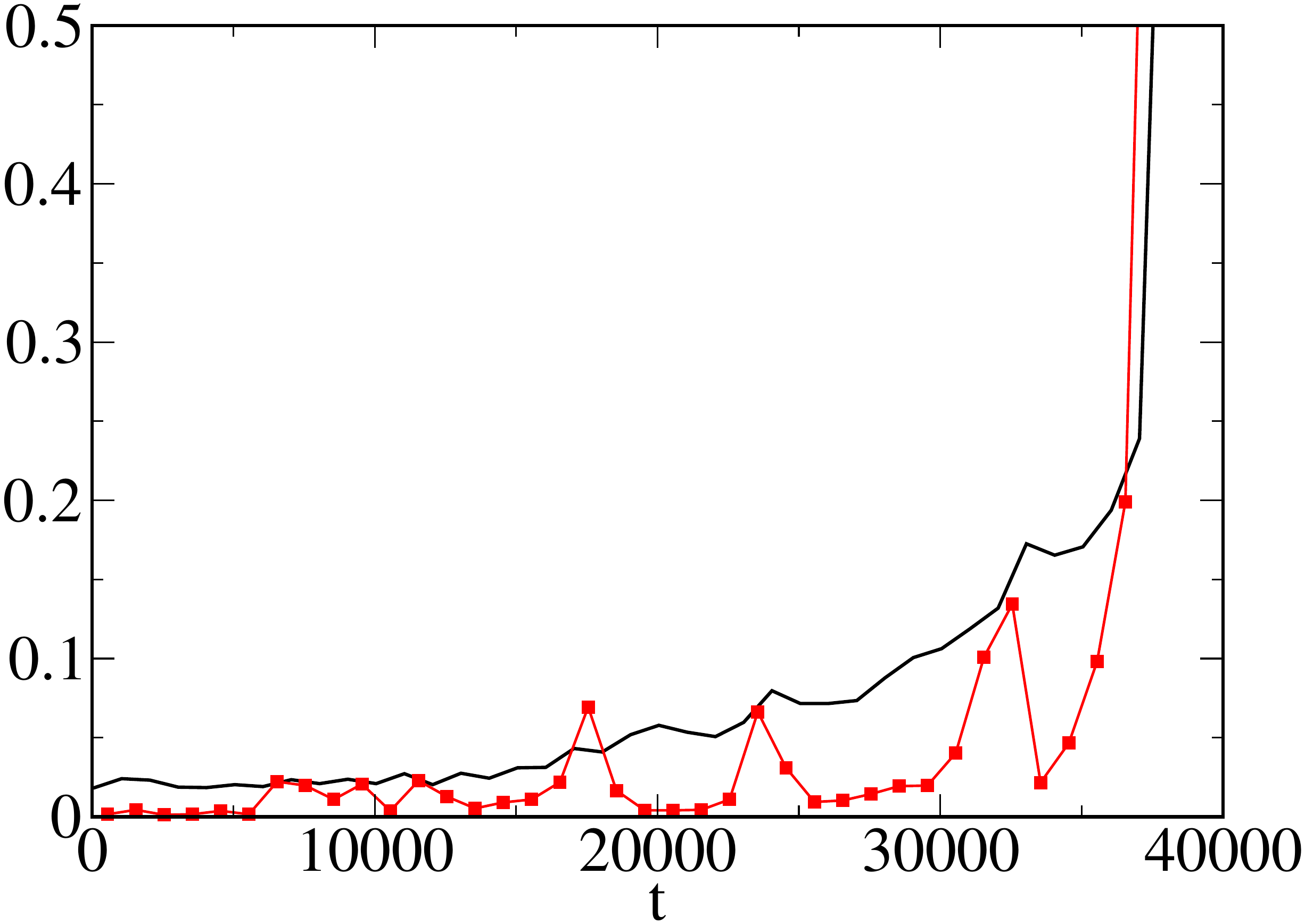}(C)
\includegraphics[width=0.45\textwidth,clip=]{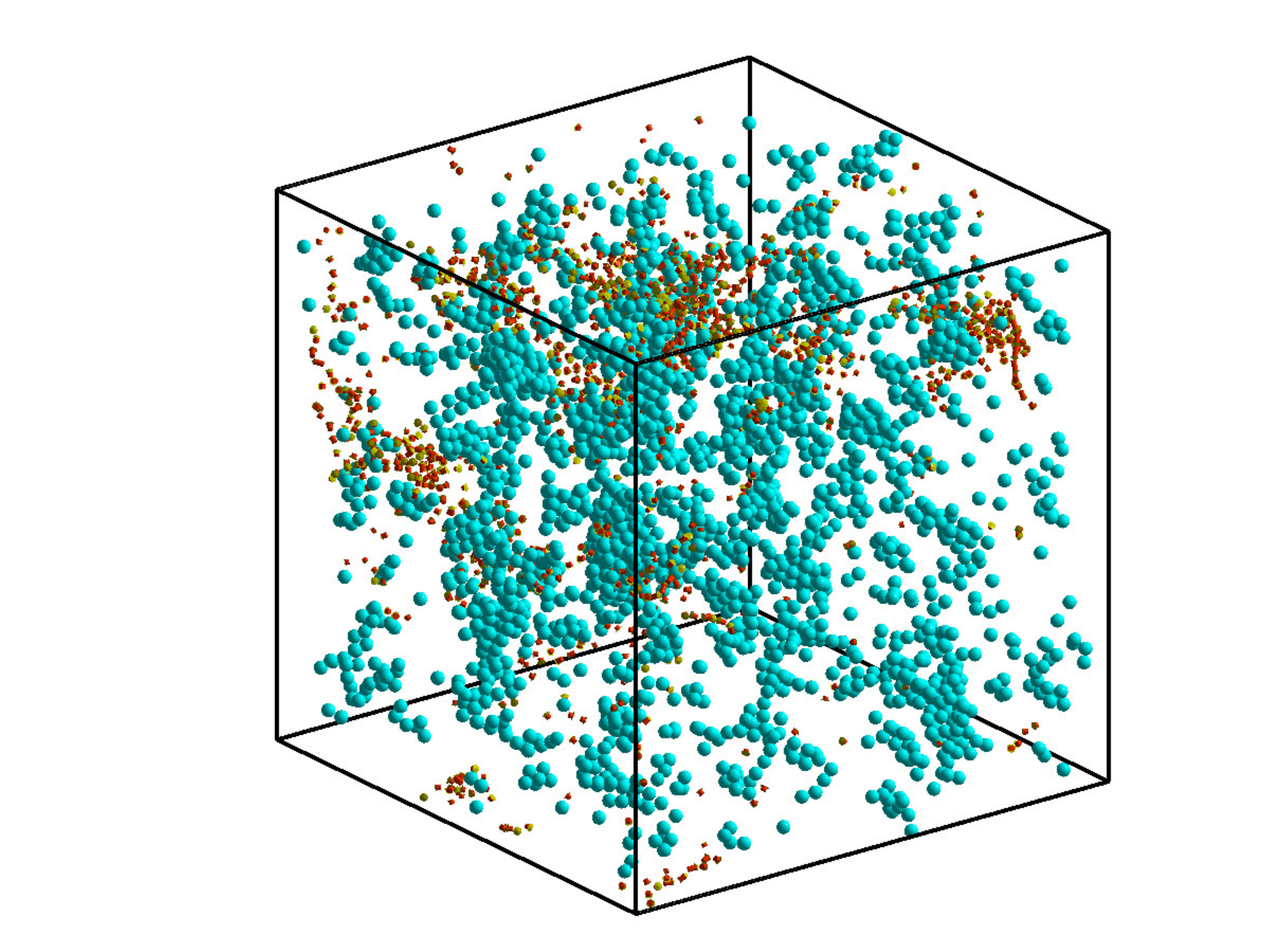}(D)
\caption{\label{sizepreparation} 
Fraction of crystalline particles and of particles belonging to an avalanche as a function of time
for different system sizes. Avalanche particles are defined in a time interval
given by the distance between consecutive points (which depends on the case under study). 
(A) a 3200-particle system (black trajectory in Fig. 1a of the main text); (B) a $3\times 3\times 3$ replica of the 3200-particle system; (C) 
a cubic subset of system (B) containing $\sim 3200$ particles; (D) Snapshot of the large system (B) at  $t=22000t_0$. 
  Solid-like particles are turquoise spheres and  avalanche particles in $[t,(t+1000t_0)]$ are red arrows with yellow
heads.}
\end{figure}

To check that our description of crystallization mechanism of an HS glass also
applies for larger systems, we initiated a run from a large configuration made
from tiling together $3\times 3 \times 3$ copies of the configuration 
used as a starting point for the trajectories shown in Fig. 1a of the main
text.  Previous work \cite{sanzPRL} shows that the artificial periodicity
induced by such spatial replication is soon lost under the randomizing
influences of the momenta (which are assigned independently in each sub-box).
Visual inspection shows that avalanches appear throughout the large system
(Fig. \ref{sizepreparation}D)).

\begin{figure}[h!]
\centering
\includegraphics[width=0.44\textwidth,clip=]{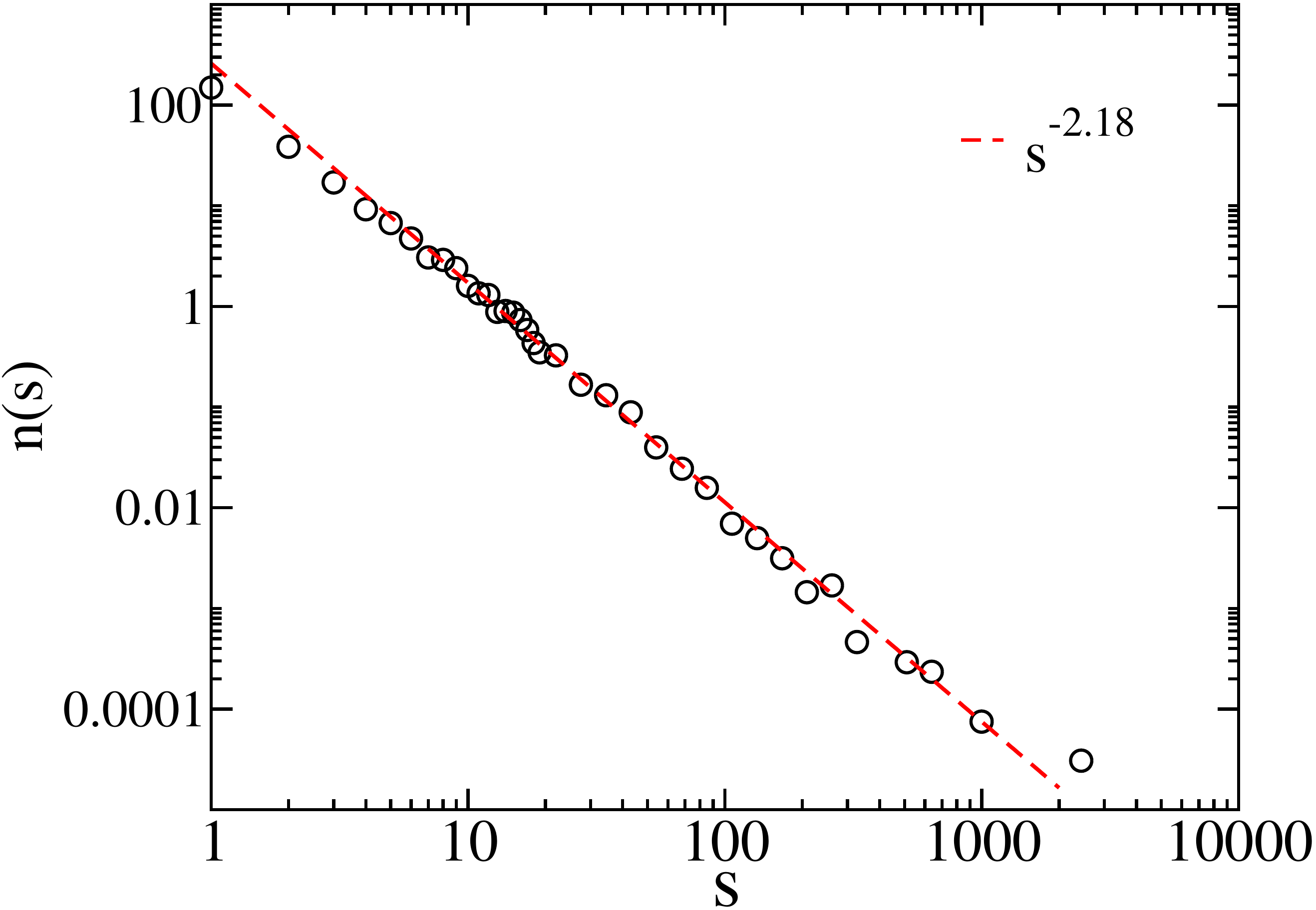}
\caption{\label{sizedistribution} 
Distribution of the size, $s$, of clusters formed by avalanche particles averaged for $X<0.1$ (black circles).
This is calculated for the large system made
from tiling together $3\times 3 \times 3$ copies of the system used in Fig. 1a of the main
text.  
The dashed red line has a slope of $-2.18$ in a log-log plot, which is the expected slope for a random percolation 
behavior.}
\end{figure}

Important differences can be seen with respect to the small system in the time
evolution of the overall crystallinity and the fraction of avalanche particles.
The first avalanche occurs sooner in the large system, as expected for a rare
event initiated by local stochasticity, and because the feedback between
avalanches and free volume is global, $X(t)$ accelerates faster thereafter.
Moreover, most of the time there is at least one avalanche present so that the
globally averaged fraction of avalanche particles, and with it $X(t)$, evolves
much more smoothly (Fig. 2B) than in the smaller systems reported above (Fig.
2A). 

On the other hand, if attention is restricted to a part of the large system
(Fig. 2C) (matched in size to the smaller systems of Fig. 2A) then the dynamics
of individual avalanches, including their extent and consequences for
crystallization, remain qualitatively similar to before 
(the dynamics of supercooled fluids shows a similar system 
size dependence \cite{appignanesiPRE2009}). Since our mechanistic
interpretation of the devitrification process is formulated at the mesoscopic
scales already captured by the simulations of 3200 particles, this
interpretation remains unaltered. Any further system-size-dependence of the
crystallization time is not expected once the density of avalanches is higher
than one per simulation box volume, which is the case in our large system.

The simulation of the replicated system allows us calculate a distribution of the size
of clusters formed by avalanche particles. A cut-off distance of 1.1 particle diameters
is used to identify neighbors in the same cluster and avalanche particles are defined
in a time interval of $500 t_0$. The cluster size (number of particles) distribution
is plotted in Fig. \ref{sizedistribution}. 
Clusters as large as $\sim 1000$ particles are observed. The distribution of cluster
sizes is typical of a random percolation, where clusters randomly appear and merge; 
this is shown by the $-2.18$ slope \cite{stauffer} of the cluster size distribution in the log-log 
plot of  Fig. \ref{sizedistribution}.
By contrast, the size distribution of crystalline
clusters found in a fresh glass at the same density 
has a slope of $-1.7$, indicating a loss of randomness
due to the preferable appearance of crystalline particles in the vicinity of
existing clusters \cite{HSclusters}. 

We note that the $X(t)$ profile of the replicated system is qualitatively similar
to that of the immature (fresh) glass investigated in Ref. \cite{sanzPRL} (Fig. 1a). 
Our preliminary investigations on this respect suggest
that avalanches are also present in  
samples of fresh monodisperse hard spheres glasses, although in a less evident form.
This would be consistent with the stochastic growth of crystals via micro-nucleation events described 
in Ref. \cite{sanzPRL}.  

\subsection{Preparation protocol of the initial configuration}

\begin{figure}[h!]
\centering
\includegraphics[width=0.44\textwidth,clip=]{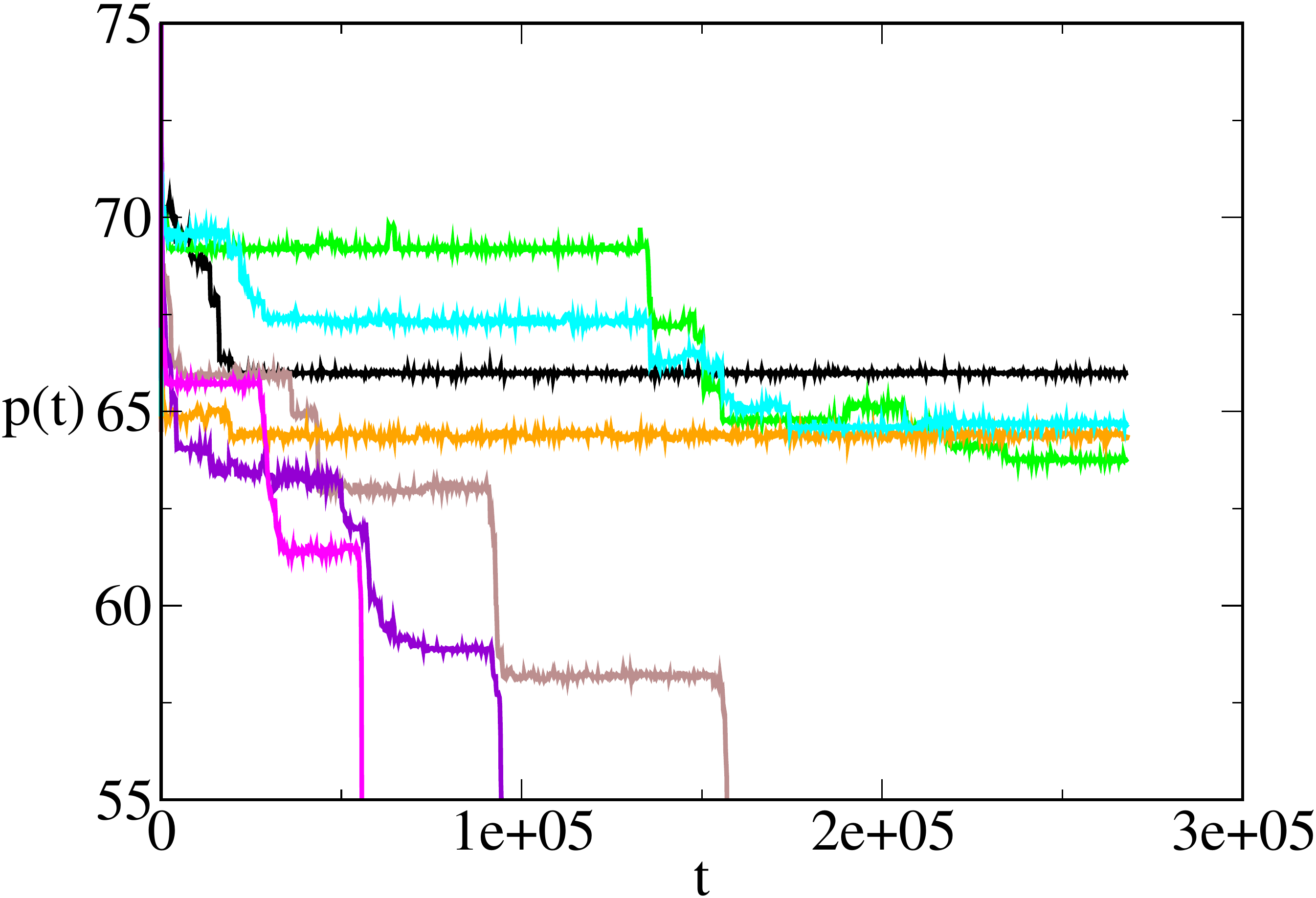}(A)
\includegraphics[width=0.47\textwidth,clip=]{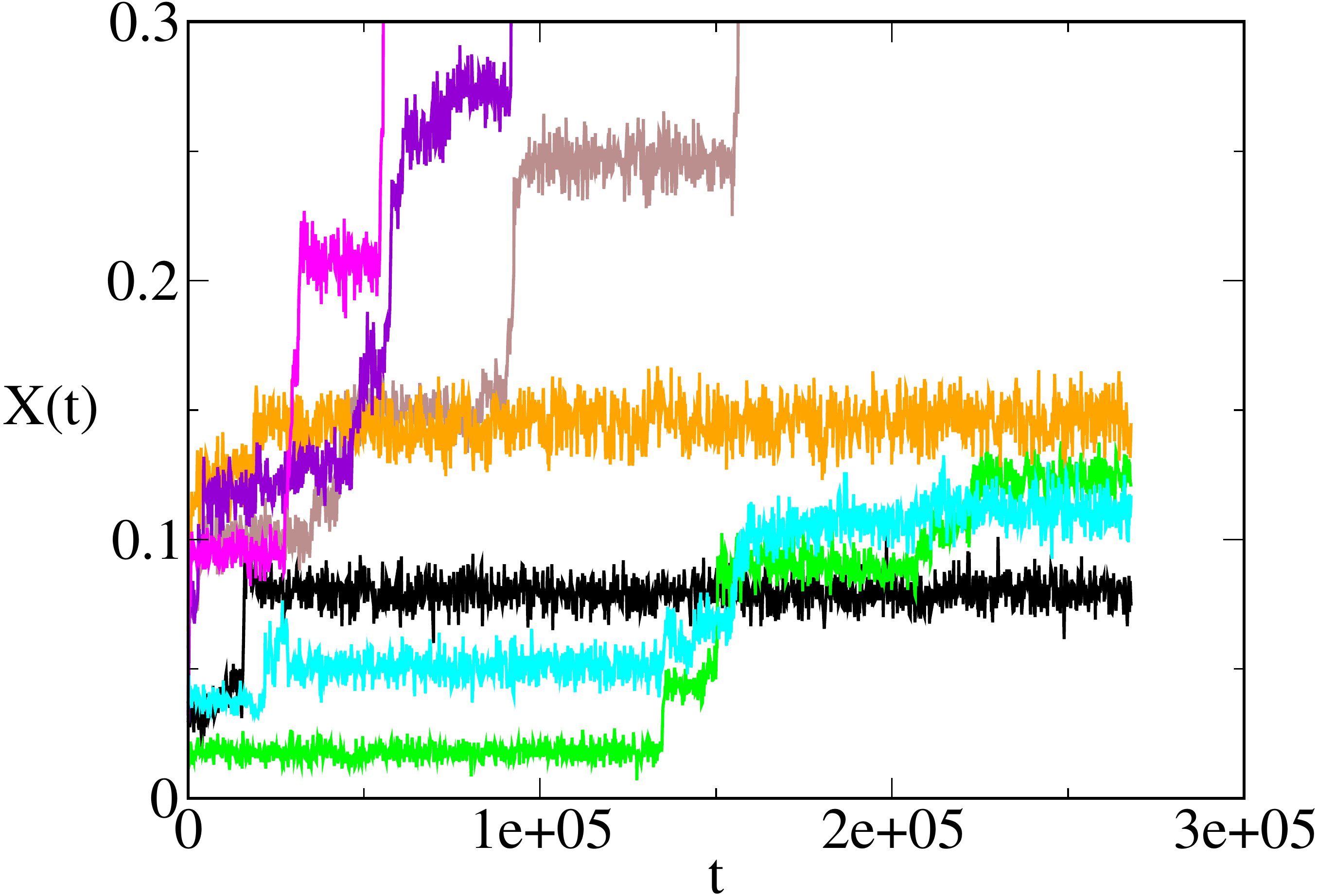}(B)
\includegraphics[width=0.44\textwidth,clip=]{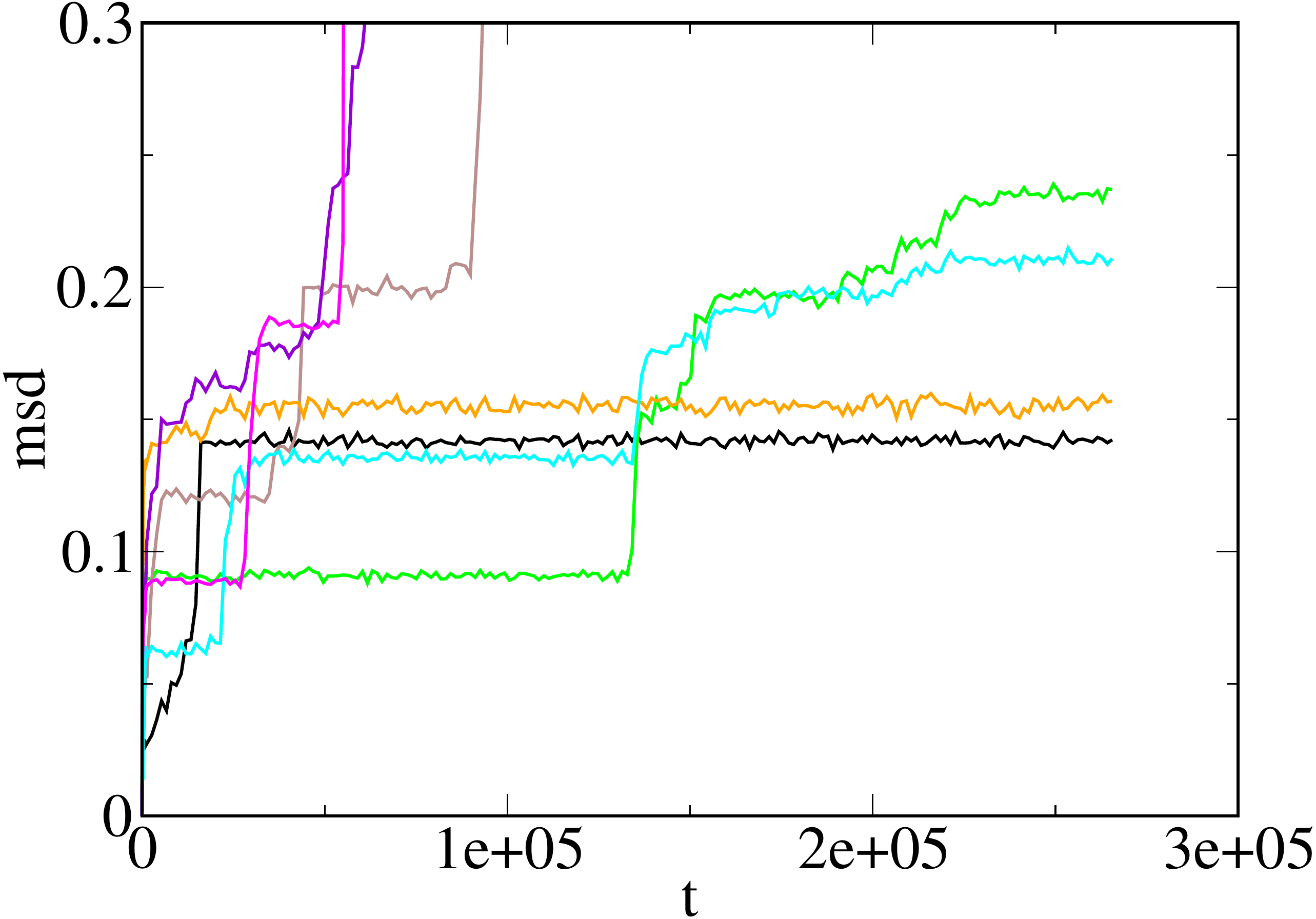}(C)
\caption{\label{phi062}Pressure (A), crystallinity $X$ (B) and msd (C) versus time for 7 independent trajectories of a 
monodisperse hard spheres suspension at $\phi=0.62$ generated by quick compression.}
\end{figure}
%We now discuss the preparation protocol of the mature glass.
%In order to avoid crystallization, we make use of the constrained ageing technique.

The results discussed in the main text correspond to glassy configurations generated with a constrained aging 
algorithm \cite{constrained} that prevents the appearance of crystallites as the system is compressed to its
final density. In this section we show that the crystallization mechanism described in the main text 
does not depend on the use of this particular protocol to generate the initial configuration.
Simply by quickly compressing the system it is also possible, although less likely, 
to obtain dense amorphous configurations of monodisperse hard spheres 
that do not readily crystallize. The odds to successfully generate such configurations
increase with the compressing rate and the target density.
We have been able to generate by quick compression configurations at $\phi=0.62$
that stay amorphous for a few decades  before crystallizing. 
In Fig. \ref{phi062} we show the time evolution of the pressure, the crystallinity and the mean squared
displacement for 7 of these configurations.
This plot is not qualitatively different from that of Fig. 1 in the main text.
Crystallization jumps are correlated to jumps in the msd (avalanches). Moreover, we also show that the pressure
drops in a sequence of steps, as a consequence of the more efficient packing achieved in crystallization events.

\subsection{Crystallization and avalanche propensity}

\begin{figure}[h]
\centering
\includegraphics[width=0.95\textwidth,clip=]{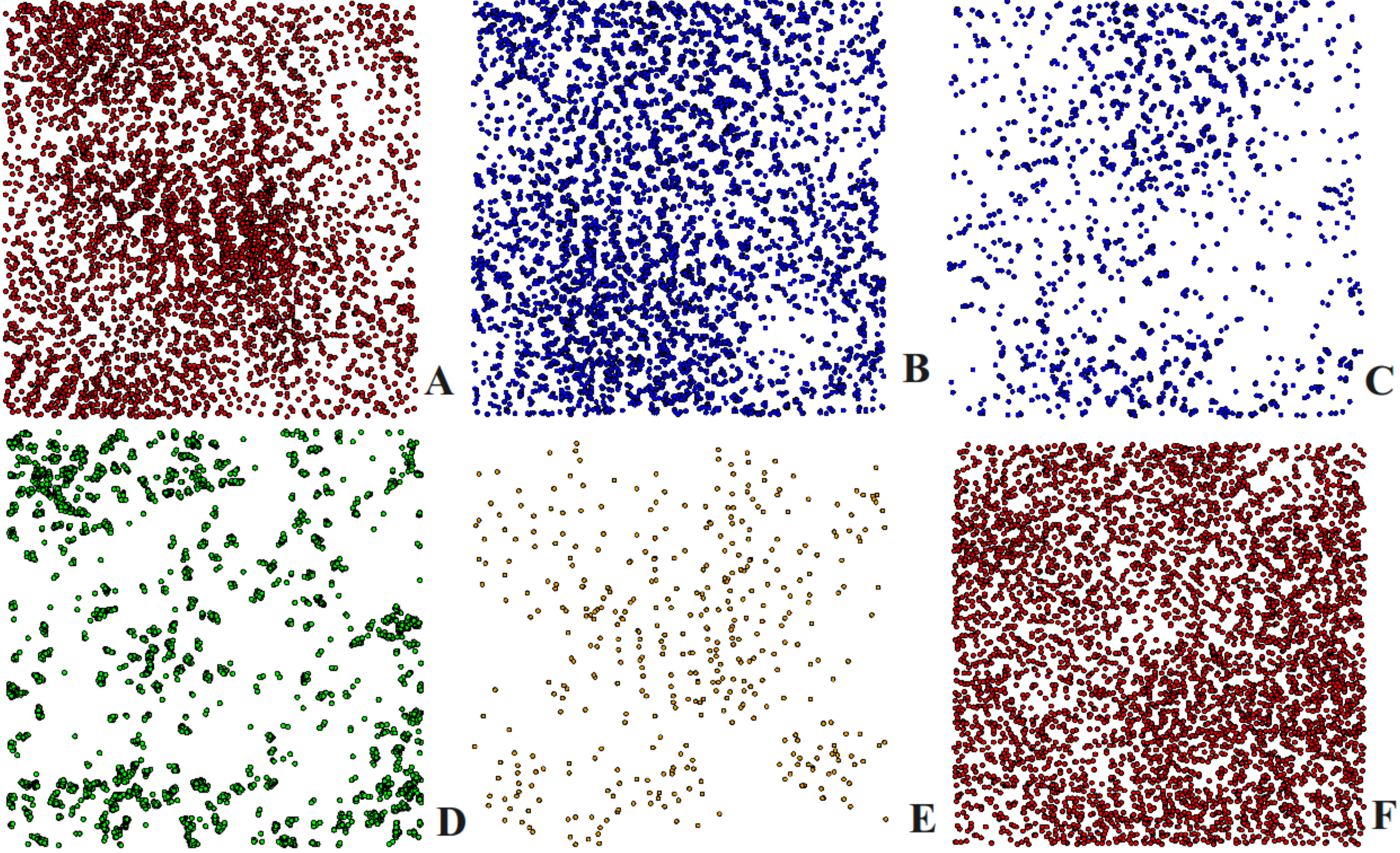}
\caption{\label{fig:XP}
A-D superimposition of different types of particles for 15 trajectories 
starting with different set of momenta from the configuration analysed in the main text:   
(A) crystalline particles at $X=0.1$ (XP);  
(B) particles involved in  the first avalanche (AP); 
(C) particles involved in the initiation of the first avalanche (AIP);
(D) particles with the top 10\% variance with respect to their average position during the initial quiescent plateau (RP). 
(E) particles with top 10\% value of the averaged local bond order parameter $\bar{q}_6$ 
 in the configuration from which all trajectories were started (MRCO).
(F) superimposed crystalline particles at $X=0.1$ of 15 trajectories starting from different configurations. 
To better observe the density distribution, the particles' size has been reduced to 30\% of their original size.
%(C) $S(q)$ computed for (A), black, and for (B), red.
}
\end{figure}

In Fig.~\ref{fig:XP} we show some snapshots to give a qualitative view of the propensities analysis
presented in the main text. 
In Fig.~\ref{fig:XP}A we show a snapshot resulting from the superimposition of the crystalline  
particles found at $X=0.1$
for the 15 trajectories of Fig.1a (main text).  
This is compared with a superimposition of the crystalline particles found at $X=0.1$
for 15 trajectories starting from different configurations (Fig.
\ref{fig:XP}F). In Fig. \ref{fig:XP}F particles are noticeably more homogeneously
distributed than in Fig. \ref{fig:XP}A, which indicates that in the
configuration from which the 15 runs of Fig.1a (main text) were initiated there
are some regions which are more prone to crystallize than others. 
Figure~\ref{fig:XP}E shows MRCO particles (see main text)
in the initial configuration of the 15 trajectories of Fig.1a. 
Comparison of Fig. \ref{fig:XP}E with Fig. \ref{fig:XP}A shows some 
correlation between MRCO regions and those with a high propensity to crystallize, 
for example a high density of particles in the centre of the simulation box. 
Figure~\ref{fig:XP}B  shows a superimposition of the particles taking part of the first avalanche (AP) 
in each of the 15 trajectories
of Fig. 1a. Differently from Fig. \ref{fig:XP}A, where there is a noticeable heterogeneity in 
the propensity to crystallize, here the probability to participate in an avalanche looks rather homogeneous throughout the system 
(density heterogeneities are comparable to those seen in the random case shown in Fig. \ref{fig:XP}F).
By contrast, when we plot in Fig. \ref{fig:XP}C only those particles involved in the initiation of the first avalanche (see below)
of each trajectory
there is a clear heterogeneity of the distribution of such particles throughout the system.  
Fig. \ref{fig:XP}D shows the superimposition of the particles with the top 10\%
variance with respect to their average position 
during the initial quiescent plateau for the 15 trajectories (rattler particles (RP)). 
We note that there is a mild {\it anti}-correlation between
MRCO and RP, as it has been
found in supercooled liquids \cite{tanaka1,tanaka2,tanaka3} (see also Fig. 8).  

%This is
%further confirmed by looking at the structure factor, $S(q)$, calculated
%for the configurations of superimposed crystalline particles. The low-q peak,
%characteristic of large-scale density heterogeneities, is much more pronounced
%for the superimposition of trajectories starting from the same configuration
%than for that obtained from trajectories starting from different
%configurations. 

%FIGURA 8

\begin{figure}[h!]
\includegraphics[width=0.44\textwidth,clip=]{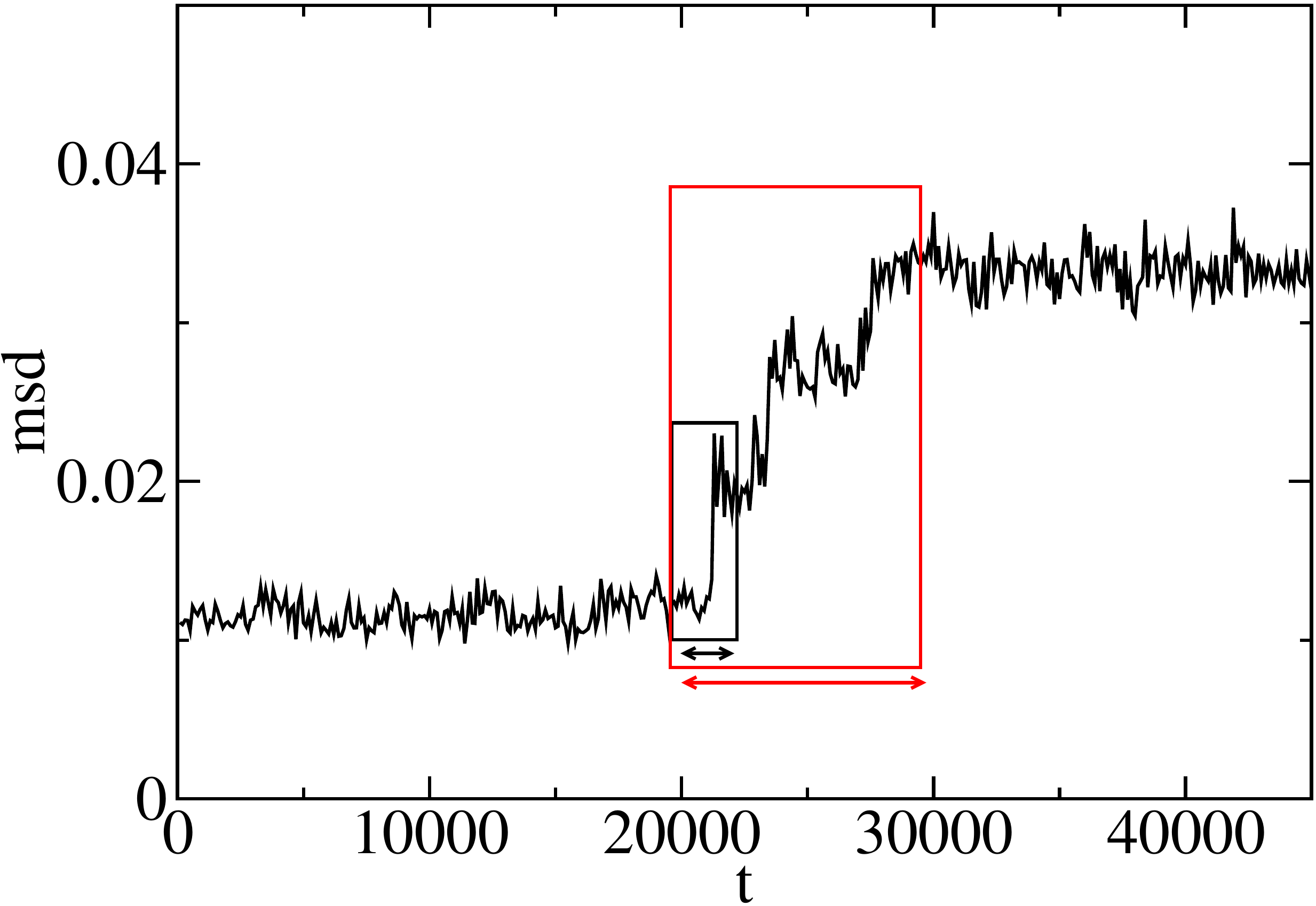}
\caption{\label{cascade} Mean squared displacement versus time for one trajectory. The avalanche is indicated 
with a red square and the time interval in which 
the avalanche is defined is given by the red arrow. Whereas the avalanche initiation is indicated with a black square 
and the avalanche initiation period is given by the black arrow.}
\end{figure}

\begin{figure}[h!]
\includegraphics[width=0.44\textwidth,clip=]{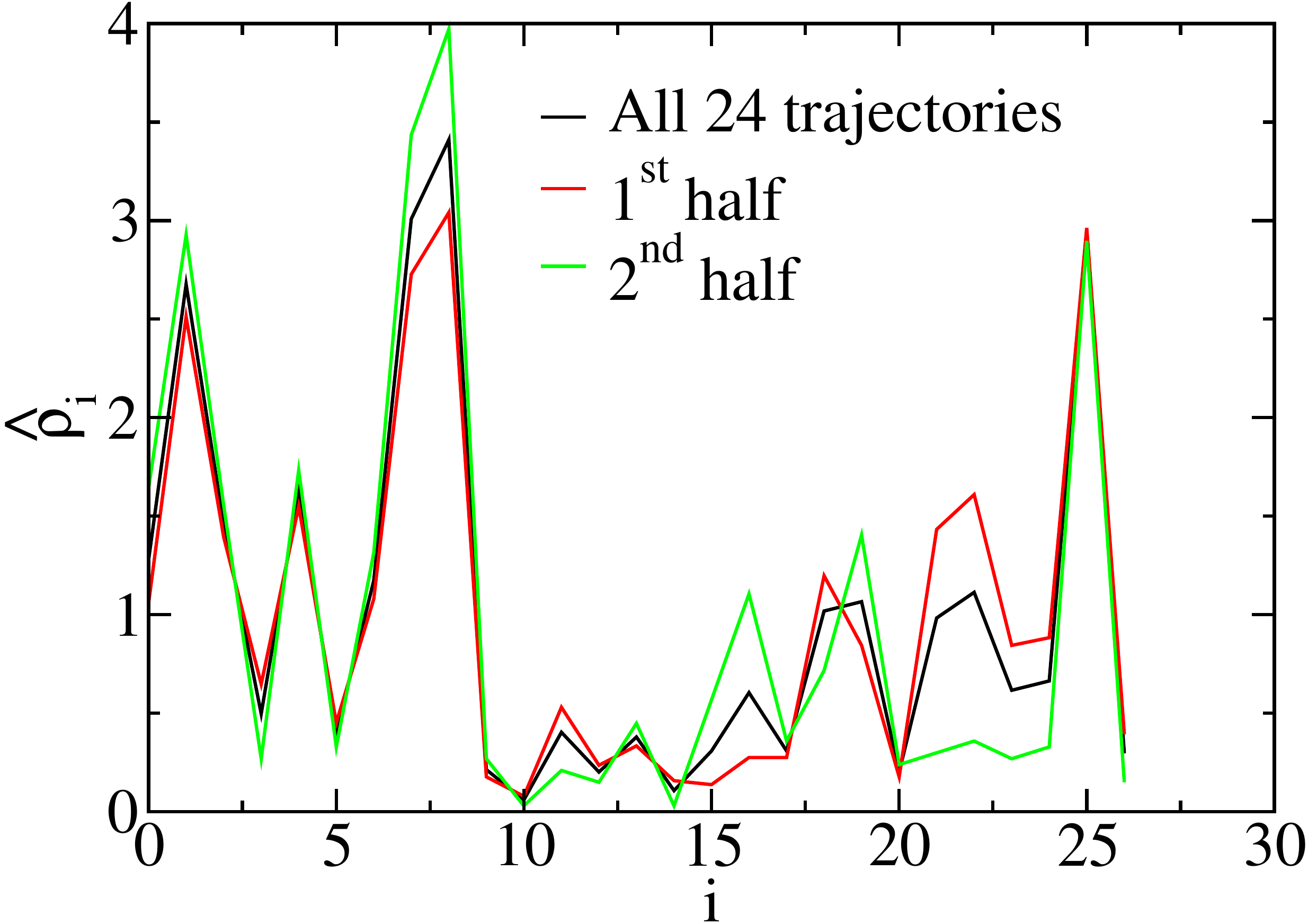}(a)
\includegraphics[width=0.44\textwidth,clip=]{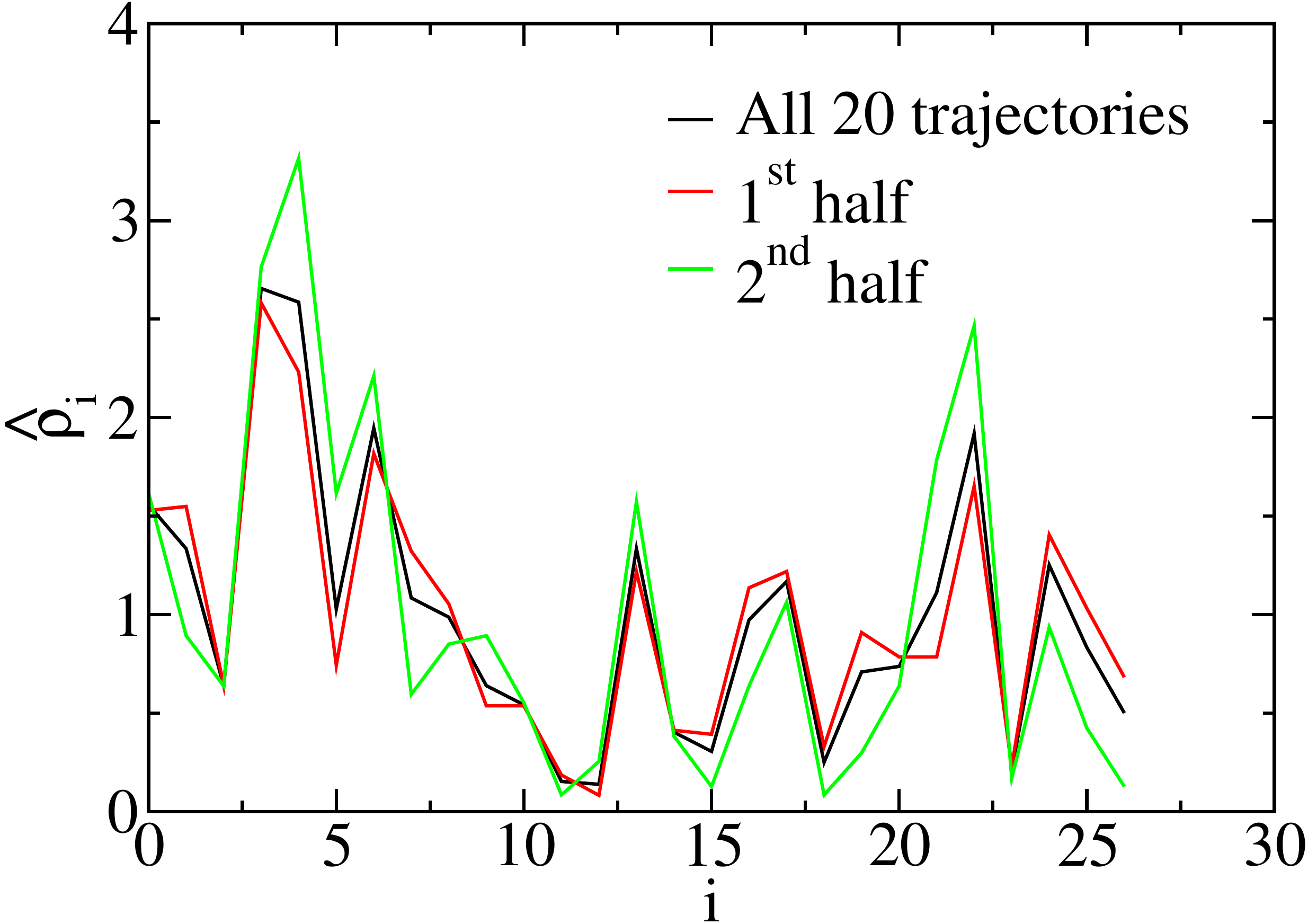} (b)
\caption{\label{statsig} Propensity curves of AIP using all available trajectories (black) and two different halves of them (red and green)
to perform the analysis. Plots (a) and (b) correspond to two different starting configurations.
Note that plot (a) corresponds to the same configuration as that analysed in Fig. 1 of the main text.}
\end{figure}

We have shown that crystallization in hard-sphere glasses tends to take place
in regions which have a high degree of medium range crystalline order.  This
observation agrees with what has been found in a number of other systems in
their more mobile ``supercooled" liquid states, e.g. \cite{tanaka2}.  However we
also found that the avalanche participants are almost randomly distributed through the
sample (green solid line in Fig. 5 (b) of the main text). This, in principle, 
suggests that there may be distinct differences between the avalanches
found in high-concentration glasses and the dynamic heterogeneity (DH) of
lower-concentration supercooled liquids (in supercooled liquids, dynamic
heterogeneities tend to develop with a higher probability in  
so-called soft spots \cite{harrowellnatphys}). However, closer inspection 
of our data reveals that avalanches are {\em initiated} preferentially in certain
regions of the system. 
As explained in the main text, we identify avalanches
by pinpointing the particles that displace beyond a certain threshold during a time interval of activity that separates two long quiescent plateaux in the mean squared displacement.
An example of such time interval is shown by the red box in Fig. \ref{cascade}.  
Careful inspection of the avalanche shown in Fig. \ref{cascade}
reveals that it 
develops as a cascade of successive mini-avalanches separated by short-lived plateaux. This 
feature is shared by most avalanches we observe. 
We therefore define avalanche initiating particles (AIP) as those that move beyond $\sigma/3$ in a time
interval that comprises {\em only the first} mini-avalanche (black box in Fig. \ref{cascade}).   
By superimposing the AIP
of 20 trajectories starting from the same
configuration -- in fact, the configuration of Fig. 1 of the main text -- with different sets of momenta we find that the propensity for an avalanche to be {\em initiated} (as opposed to avalanche {\em participation}) is clearly heterogeneously
distributed  
 (light-green dashed line in Fig. 5 (b) of the main text). 
Fig. \ref{statsig}, where the AIP propensity curves of two different halves of the total available trajectories are compared, 
shows that this result is statistically significant. 
Therefore, avalanches, like DH, tend 
to be triggered preferentially in particular regions of the system. 

\begin{figure}[h!]
\includegraphics[width=0.44\textwidth,clip=]{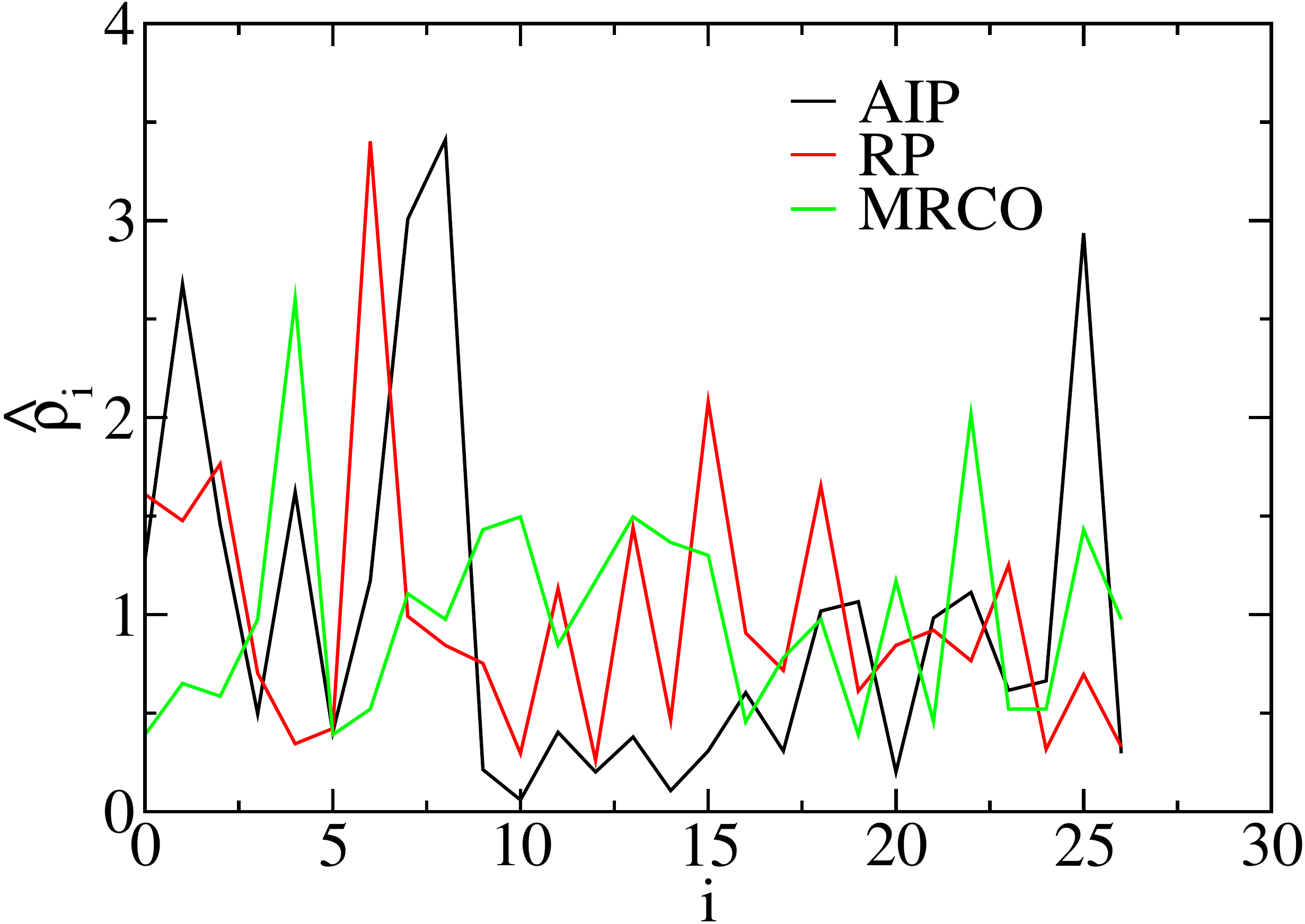} (a)
\includegraphics[width=0.44\textwidth,clip=]{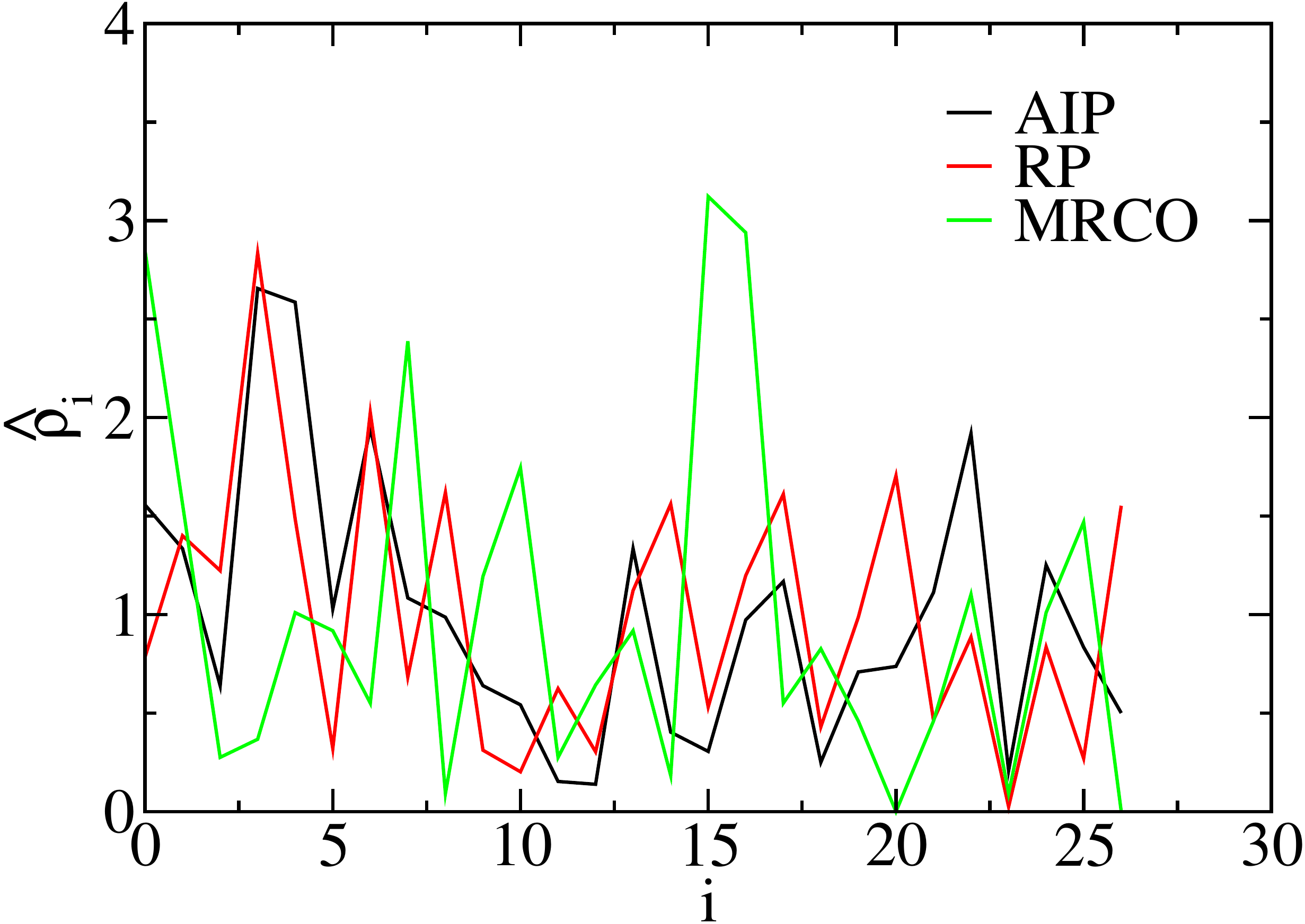} (b)
\caption{\label{avavsratt} Normalised density for the 27 sub-volumes in which the system is divided 
for different types of particles (as indicated in the legend) and 
for two different configurations (a) and (b).
Plot (a) corresponds to the same configuration as that analysed in Fig. 1 of the main text.}
\end{figure}

To investigate further any connection between avalanches and DH, we next
inspect whether AIP-rich regions are also rich in particles that have a high
rattling freedom in the initial quiescent plateau. (Such rattlers are in turn
equivalent to soft spots, which were found to be correlated to DH in metastable
fluids \cite{britowyart,harrowellnatphys,harrowell2009,yodh}). 
 In Fig.
\ref{avavsratt} we show the density profile of AIP particles compared to that
of rattling particles (RP) for two set of simulations started from two different configurations. Our results are
not conclusive, but offer plausible evidence of some correlation in at least
one of these two configurations. If this is confirmed by future work,
avalanches could plausibly be viewed as a limiting type of DH that arises when
activity becomes rare as the system's density/age increases. In table \ref{summava} 
we summarize the comparison between the characteristics of 
avalanches in a crystallizing hard sphere glass and those of
dynamic heterogeneities in a supercooled fluid. 

\begin{table}
\footnotesize
\begin{tabular}{c c c c }
 \hline
 \hline
Property  & DH & A  \\
\hline
Heterogeneous in space & Yes \cite{nmr,kob_dh}  & Yes   \\ 
Cooperative dynamics   & Yes \cite{donatipoole}  & Yes   \\ 
Stochastic in space and time             & Yes \cite{harrowellrandom,d-clusters2006}  & Yes  \\ 
Spacial propensity  & Yes \cite{harrowellrandom} & Yes (initiation)   \\ 
Propensity correlated to soft spots  & Yes \cite{harrowellnatphys} & Further investigation required  \\ 
\hline
 \hline
\end{tabular}
\caption{Comparison between the properties of dynamic heterogeneities (DH) in supercooled fluids and  
avalanches (A) in a crystalizing hard sphere glass (this work).}
\label{summava}

\end{table}

%FIGURA 10
%\begin{figure}[h]
%\centering
%\includegraphics[width=0.2\textwidth,clip=]{./FIGS/mrco.eps}
%\caption{\label{fig:MRCO} MRCO regions in the configuration from which the 15 trajectories with independent momenta ($R1,R2,...,R15$) were started.}
%\end{figure}

%\begin{figure}[h]
%\centering
%\includegraphics[width=0.24\textwidth,clip=]{./FIGS/densitycorrelations.ok2.eps}\includegraphics[width=0.2%4\textwidth,clip=]{./FIGS/densitycorrelations.ok3.eps}

%\includegraphics[width=0.24\textwidth,clip=]{./FIGS/densitycorrelations.ok4.eps}
%\includegraphics[width=0.23\textwidth,clip=]{./FIGS/compareboxesavalancheandotherstaff_AP_DP.eps}

%\includegraphics[width=0.23\textwidth,clip=]{./FIGS/compareboxesavalancheandotherstaff_AP_XP.eps}
%\includegraphics[width=0.23\textwidth,clip=]{./FIGS/compareboxesavalancheandotherstaff_AP_MRCO.eps}
%\caption{\label{fig:denscorr} Normalised density, $\rho_i/\rho_{total}$, as a function of the index $i$ identifying each sub-box. 
%The normalised density is calculated for $XP$, $MRCO$, $AP$ and $DP$ (from trajectory $R10$) particles %and compared as indicated in the legend.}
%\end{figure}

%{\bf XXX from page 10 in the text  (POSSIBLY COMMENT IN SUPP MAT ON DYNAMIC PROPENSITY AND ANTI-CORRELATION WITH MRCO) ?}

%Likewise, the crystallization propensity is assessed by superimposing the solid-particles of all 
%trajectories as these first cross a fixed crystallinity threshold (we choose $X=0.1$). (PUT THESE PICTURES IN SUPP MAT ?)

\subsection{Online video}

 The video (Movie SI) represents solid and avalanche particles participating to the avalanche shown in Fig.~1(b) of the main text.
Solid-like particles are turquoise spheres and  avalanche particles in $[t,(t+1000t_0)]$ are red arrows with yellow
heads.
 The avalanche starts to build in localized regions, 
 then grows to peak activity, and finally dies out leaving behind an increased population of solid-like particles. 
Highly cooperative movements can be seen during the main avalanche phase, including particles
moving in rows or circles.  
 From start to finish, an avalanche typically lasts about $7000t_0$, with $t_0$ the time unit defined in the Methods Section.

%\end{article}

\end{document}